\documentclass[aps,twocolumn,superscriptaddress,a4paper]{revtex4}
\usepackage{amsfonts,amssymb,amsmath}
\usepackage[utf8]{inputenc}
\usepackage[pdftex,breaklinks=true,colorlinks=true]{hyperref}
\usepackage[pdftex]{graphicx}
\usepackage{epstopdf}
\usepackage{leftidx}
\graphicspath{{Pictures/}}

\begin{document}

\title{Out-of-equilibrium dynamics in a quantum impurity model: Numerics for particle
transport and entanglement entropy
}

\author{Kemal Bidzhiev}
\affiliation{Institut de Physique Th\'eorique, Universit\'e Paris Saclay, CEA, CNRS, F-91191 Gif-sur-Yvette, France}
\author{Gr\'egoire Misguich}
\affiliation{Institut de Physique Th\'eorique, Universit\'e Paris Saclay, CEA, CNRS, F-91191 Gif-sur-Yvette, France}

\date{\today}

\begin{abstract}
We investigate the out-of-equilibrium properties of a simple quantum impurity model, the interacting resonant level model. We focus on the scaling regime, where the bandwidth of the fermions in the leads is larger than all the other energies, so that the lattice and the continuum versions of the model become  equivalent. Using time-dependent density matrix renormalization group simulations initialized
with states having different densities in the two leads
we extend the results of Boulat, Saleur and Schmitteckert [Phys. Rev. Lett. 101, 140601 (2008)] concerning the current-voltage ($I$-$V$) curves, for several values of the interaction strength $U$. We estimate numerically the Kondo scale $T_B$ and the  exponent $b(U)$ associated to  the tunneling of the fermions from the leads to the dot. Next we analyze the quantum entanglement properties of the steady states. 
We focus in particular on the entropy rate $\alpha$, describing the linear growth with time of the bipartite entanglement in the system. We show that, as for the current, $\alpha/T_B$ is described by some function of $U$ and of the rescaled bias $V/T_B$. Finally, the spatial structure of the entropy profiles is discussed.
\end{abstract}
\maketitle
\section{Introduction}

The dynamical properties of out-of-equilibrium quantum many-body systems are a major topic in condensed-matter physics~\cite{Polkovnikov_RMP,eisert_quantum_2015}.
Although the equilibrium properties of one-dimensional (1d) interacting problems are well understood in many cases, thanks, for instance, to some theoretical techniques such as the Bethe Ansatz, conformal field theory or matrix-product states (MPS) numerics~\cite{schollwock_density-matrix_2011}, the physics that takes place in out-of-equilibrium situations represents very rich domain, and is much less understood.

Quantum impurity problems are among the simplest quantum many-body systems, but they nevertheless harbor many interesting phenomena, many open questions, and they represent a very useful playground to develop new theoretical ideas and methods. They are also -- of course -- relevant to describe many experimental situations, from the Kondo effect in metals \cite{hewson_kondo_1993} to transport in nanostructures such as quantum dots or point contacts.

In this work we consider a well known impurity model - the interacting resonant level model (IRLM) \cite{wiegmann_resonant-level_1978}. The model describes two semi-infinite leads with spinless fermions that are coupled to some resonant level (dot) via some tunneling and some interaction [see Eq.~(\ref{eq:H_IRLM}) below].
We are interested here in the transport properties of the system, and wish to describe quantitatively the
steady states that appear when some particle current is flowing through the dot. How do we address these questions, without relying on
the linear response theory, nor  using some perturbative scheme that would assume that the system is close to equilibrium and/or that the interactions are weak ?
Thanks to the integrability of the IRLM~\cite{filyov_method_1980}, several remarkable exact results have been obtained concerning non-equilibrium steady states, 
such as the current-voltage ($I$-$V$) characteristic for some special (so-called ``self-dual'') value of the interaction strength \cite{boulat_twofold_2008} (see also \cite{fendley_exact_1995,fendley_exact_1995B,mehta_nonequilibrium_2006}).
However, apart from  this special point and the noninteracting case,  simple quantities such as the $I$-$V$ curve are not known analytically.
From this point of view numerical simulations are invaluable and complementary to the analytical approaches.

This study indeed aims at  providing accurate numerical data concerning the  so-called scaling regime of the lattice model
({\it i.e.} all the energies are small compared to the bandwidth in the leads),  where many quantities become universal and can be quantitatively compared to 
the field theory results (continuum limit).

To simulate transport, a useful setup is to prepare a large but finite isolated system in an initial state, at $t=0$, where two spatial regions -- say the left and the right leads -- have different
particle densities. Starting from such an inhomogeneous state, the Hamiltonian evolution of the wave function will put the particles in motion. This first leads to some transient regime where some current starts flowing from one side to the other. For times that are long compared to the microscopic time scales, but
smaller than the time required for an elementary excitation to propagate through the whole system, we expect on physical grounds that some quasi-steady states should be realized.
This type of approach, where one follows numerically the real time evolution starting from an initial state with finite density bias, has already been used to investigate several
interacting 1d systems, like XXZ spin-$\frac{1}{2}$ chains~\cite{gobert_real-time_2005,sabetta_nonequilibrium_2013}, or impurity models~\cite{schmitteckert_nonequilibrium_2004,schneider_conductance_2006,boulat_twofold_2008}.
These simulations have been performed using the time-evolving block decimation and time-dependent density matrix renormalization group (DMRG) \cite{vidal_efficient_2004,white_real-time_2004,daley_time-dependent_2004}, where the many-body wave function of the system is encoded as a matrix-product state.

Our objective is first to refine and extend some of the previous numerical studies concerning the particle current in the IRLM \cite{boulat_twofold_2008}, and then to focus on the entanglement entropy.
Even though there have been several studies on the entanglement properties of quantum impurity models, most of these works focused on the ground state (for a review see \cite{laflorencie_quantum_2016}). Here we will analyze in detail the linear growth of the entanglement entropy with time, and its spatial structure in the steady regime.

The plan of the paper is as follows.
Section.~\ref{sec:model_evol} presents the model, the initial state, and describes qualitatively the evolution of three  quantities
of interest: particle density, particle current, and  entanglement (von Neumann) entropy.
The central results are then presented in Sec.~\ref{sec:ness}. This section is devoted to the steady state, and presents some quantitative analysis of the numerics for the steady current and entropy rate, focusing on the scaling regime of the model.
In particular we confirm (Sec.~\ref{ssec:sc}), following Ref.~\cite{boulat_twofold_2008}, that in this limit the current $I$ is some universal function of $V/T_B$ and $U$, where $V$ is the initial bias, $U$ the interaction strength, and $T_B$ is Kondo crossover scale that we evaluate numerically.
In Sec.~\ref{ssec:entropy_rate} we present a similar analysis for the scaling of the entropy rate, and
we show that it can also be analyzed in term of some universal functions of $V/T_B$ that we compute numerically for several values of $U$.
The shape of the out-of-equilibrium  entanglement profile is discussed in Sec.~\ref{ssec:entropy_profile}.

Finally, Appendix~\ref{sec:numerics} provides some technical details concerning the numerical simulations, and Appendix~\ref{sec:ff}
presents a few exact results in the free-fermion case (current, density and entropy rate).

\section{Model and time evolution}
\label{sec:model_evol}

\subsection{Hamiltonian}

\begin{figure}[h]
\includegraphics[scale=0.3]{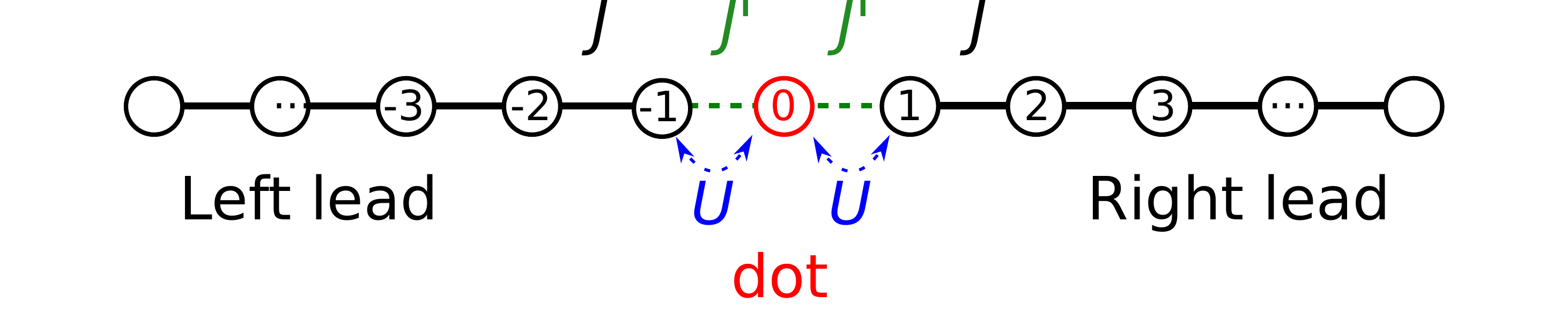}
\caption{Schematics of the interacting resonant level model. The system is prepared at $t=0$
in the ground state of the model with an additional chemical potential equal to $+V/2$ in the left lead, and equal to $-V/2$ in the right lead.
For $t>0$ the system then evolves with the bias switched to zero.
In all cases the chemical potential is zero on the dot, hence the name ``resonant''.
Note that, by symmetry, the average fermion occupation number $\langle c_0^\dagger c_0\rangle$ on the dot is equal to $\frac{1}{2}$.}
\label{fig:IRLM}
\end{figure}

We consider a lattice version of the IRLM (Fig.~\ref{fig:IRLM}), which can be defined as
\begin{eqnarray}
 &H_{\rm IRLM}&=H_A+H_B +H_d, \\
 &H_A&=-J\sum_{r=-N/2}^{-2} \left(c^\dagger_r c_{r+1} +\text{H.c}\right), \\
 &H_B&=-J\sum_{r=1}^{N/2-1} \left(c^\dagger_r c_{r+1} +\text{H.c}\right), \\
  &H_d&=-J' \sum_{r=-1}^0 \left( c^\dagger_{r} c_{r+1} +\text{H.c}\right) +\nonumber\\
  &&+ U\sum_{r=\pm 1} \left(c^\dagger_r c_r-\frac{1}{2}\right)\left(c^\dagger_0 c_0-\frac{1}{2}\right),
  \label{eq:H_IRLM} 
 \end{eqnarray}
where $H_{A/B}$ describes the kinetic energy of free spinless fermions in  the left and right leads,  and $H_d$ models
the tunneling from the leads to the dot (site at $r=0$), as well as the density-density interaction between the dot and the ends of the leads ($r=-1$ and $1$).
 
As discussed later we will focus on the regime where the  hopping amplitude  $J'$ (or  tunneling strength) between the leads and the dot ($r=0$) is much
smaller than the bandwidth $4J$ of the kinetic energy in the leads, i.e $J'\ll J$.
From now we take $J=1=\hbar$, thus defining the unit of time and energy.

\subsection{Initial state}

We choose the initial state $|\psi(t=0)\rangle$ to be the ground state of $ H_0 = H_{\rm IRLM}+H_{\rm bias}$, where
$H_{\rm bias}$ is  an inhomogeneous chemical potential (or voltage) that induces different densities on the left and on the right leads:
\begin{equation}
 H_{\rm bias} = \frac{1}{2}V \sum_{r=-N/2}^{N/2-1} \tanh(r/w) c^\dagger_r c_{r}.
 \label{eq:tanh}
\end{equation}
In the left (right) lead the chemical potential is thus equal to $V/2$ ($-V/2$) sufficiently far from  the dot.
This bias induces different initial densities $\langle c^\dagger_r c_{r}\rangle=\frac{1}{2}\pm m_0$ in the bulk of the leads at $t=0$ (blue horizontal
line in Fig.~\ref{fig:Szprofile}).
For an infinite system, $N \to \infty $, the Fermi momenta $k_F^+$ ($k_F^-$) in the left (right) lead is set by $2J\cos(k_F^\pm)=\pm V/2$,
and these are related to the density difference $m_0$  through $k_F^\pm=\pi(\pm m_0+1/2)$.

As done in previous studies \cite{boulat_twofold_2008}, the voltage drop in Eq.~(\ref{eq:tanh}) is spatially smeared over $\sim w$  sites in the vicinity of the dot. This
has the effect of producing an initial state with smoother density in the vicinity of $r=0$ and turns out to accelerate the convergence to a steady state.
We typically use $w=10$. For the same reason, the initial state is prepared with $J'=J$, that is uniform hopping amplitudes throughout the chain.
In addition, $H_0$ is  chosen to be noninteracting ($U=0$), and the interactions are switched on for $t>0$.

\subsection{Unitary evolution at $t>0$}

For $t>0$ the wave function evolves using $H_{\rm IRLM}$, with $0<J'<1$, and without the voltage bias term.
The calculations are performed using a time-dependent DMRG algorithm, implemented using the C++ iTensor library \cite{itensor}.
The evolution operator $U=\exp(-i\tau H)$ for a time step $\tau$ is approximated by a matrix-product operator (MPO) \cite{zaletel_time-evolving_2015}, using a fourth-order Trotter scheme [Eqs.(\ref{eq:trotter}) and (\ref{eq:trotter4})] and $\tau=0.2$ (unless specified otherwise). The system sizes are of the order of 200 sites and the largest times are of the order of $t\simeq 100$.
Additional details about the simulations are given in Appendix~\ref{sec:numerics}.

In the following we  focus on three quantities:
the particle density (Sec.~\ref{ssec:density}), the particle current (Sec.~\ref{ssec:current}), and the entanglement entropy (Sec.~\ref{ssec:entropy}).

\subsection{Particle density}
\label{ssec:density}

\begin{figure}[h!]
\includegraphics[height=0.5\textwidth, angle=270]{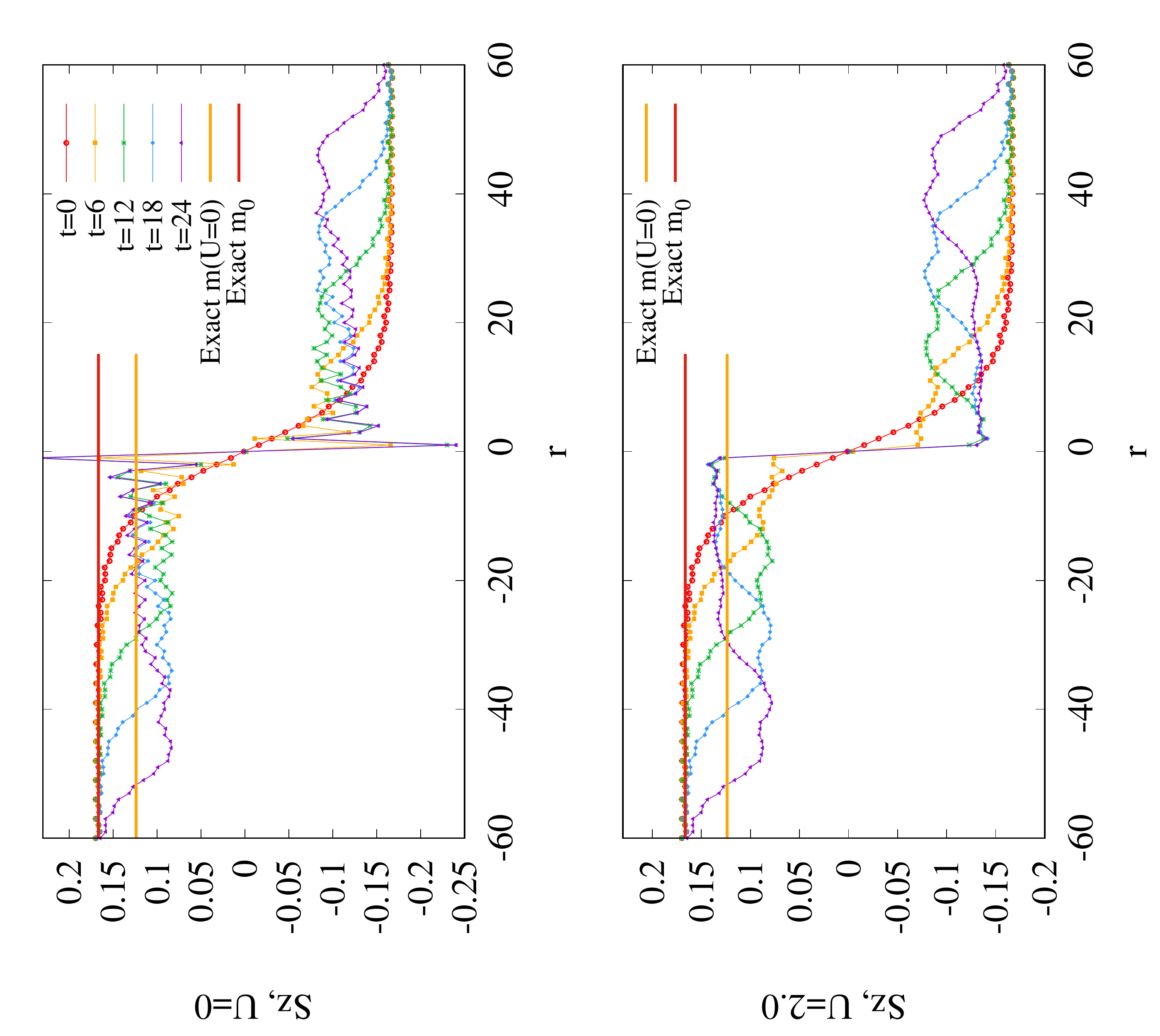}
\caption{Magnetization  profile $S^z(r,t)$ at times $t=0,6,12,18,24$. Parameters of the model: $J'=0.3$, $V=2.0$, $N=254$.
Top panel: $U=2.0$, bottom: $U=0$. In both cases the initial magnetization $m_0$ in the bulk of the left lead is indicated by a (red) horizontal line.
The orange horizontal line marks the exact bulk stationary magnetization $m$ for $U=0$ [see Eq.~(\ref{eq:m})]. The data in the upper panel show some deviations
from this non-interacting value. Note that some Friedel-like oscillations develop at long times in the vicinity of the dot, and these are much stronger in the non-interacting case.}
\label{fig:Szprofile}
\end{figure}

The particle density is defined by $\rho(r,t)=\langle \psi(t) | c^\dagger_{r} c_{r} |\psi(t)\rangle$,
and we can equivalently use a spin language, with the  magnetization $S^z(r,t)=\rho(r,t)-\frac{1}{2}$.
A typical evolution of the density profile is shown in Fig.~\ref{fig:Szprofile}. It shows
how the initial profile at $t=0$ gives rise to 
two propagating fronts (one to the left and one to the right), forming a ``light cone'', and how some steady region form in the center. When the time is large enough,
two regions with quasi homogeneous densities develop on both sides of the dot. The densities in the steady
regions of the lead can be written as $\rho=\frac{1}{2}\pm m$, and the density difference $m$ between both sides of the dot can be computed exactly in the free-fermion case $U=0$. The result reads as:
\begin{equation}
 m=\int_{k_F^-}^{k_F^+} \frac{dk}{2\pi} \mathcal{R}(\epsilon(k))
 \label{eq:m}
\end{equation}
where $\mathcal{R}(\epsilon)$ is the reflexion coefficient for an incident particle with energy $\epsilon$ (more details in Appendix.~\ref{ssec:dens_drop}).
This exact value $m$ is in agreement with the magnetization that is measured numerically in the stationary region for $U=0$ (bottom panel of Fig.~\ref{fig:Szprofile}),
but slightly different from that observed in the interacting case (top panel of Fig.~\ref{fig:Szprofile}), as expected.

\subsection{Particle current}
\label{ssec:current}
On a given bond the expectation value of the current operator is $I(r,t)=2 J_r {\rm Im} \langle \psi(t) | c^\dagger_r c_{r+1} |\psi(t)\rangle$, where $J_r$ is the hopping amplitude
between the sites located at $r$ and $r+1$. We thus have $J_{r}=J'$ for $r=-1$ and $r=0$ (hopping to the dot), and $J_{r}=J=1$ otherwise (in the leads).
This definition ensures the proper charge conservation equation, $\frac{d}{dt} \langle \psi(t) |c^\dagger_r c_r|\psi(t)\rangle =I(r-1,t)-I(r,t)$.
When no position $r$ is given, $I(t)$ refers to the averaged current on both sites of the dot: $I(t)=\frac{1}{2} \left(I(-1,t)+I(0,t)\right)$. 

\begin{figure}[h!]
\includegraphics[height=0.5\textwidth, angle=270]{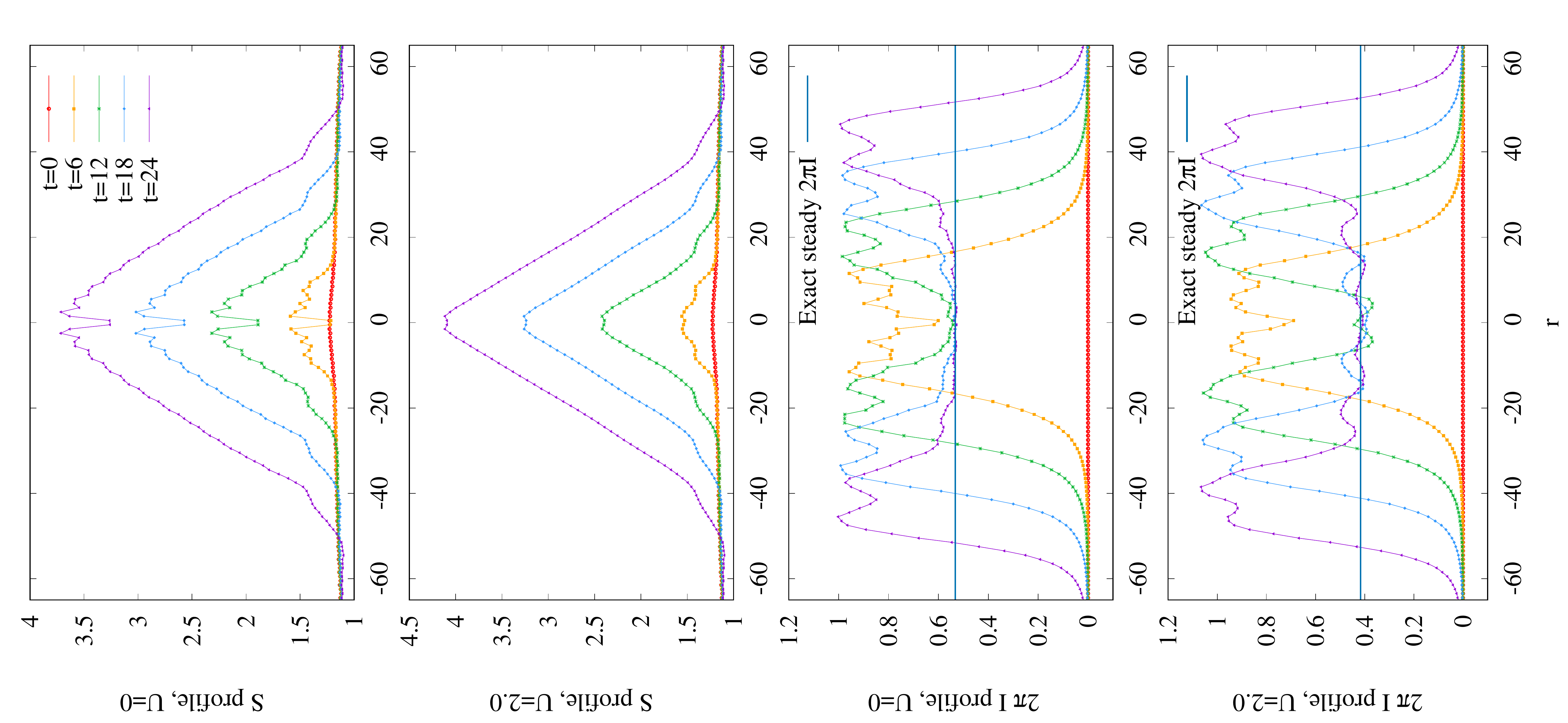}
\caption{Top panels: evolution of the entropy $S(r,t)$ profile for $U=0$ and $2.0$. Bottom panels: evolution of the current profile $I(r,t)$ for $U=0$ and $2.0$. Parameters of the model: $J'=0.3$, $V=2.0$, $N=254$, as in Fig.~\ref{fig:Szprofile}.}
\label{fig:profiles}
\end{figure}

A typical evolution of the current profile is shown in the upper panel of Fig.~\ref{fig:profiles}. As for the density, two propagating fronts
are  visible. In the center of the system one observes, at sufficiently long times, the emergence of a spatial region where the current is almost constant in space and time.
This is where some steady value of the current can be defined. Interestingly, the current in the (left-moving  and right-moving) front regions reaches values that are significantly larger than in the steady regions.

\begin{figure}[h]
\includegraphics[height=0.5\textwidth, angle=270]{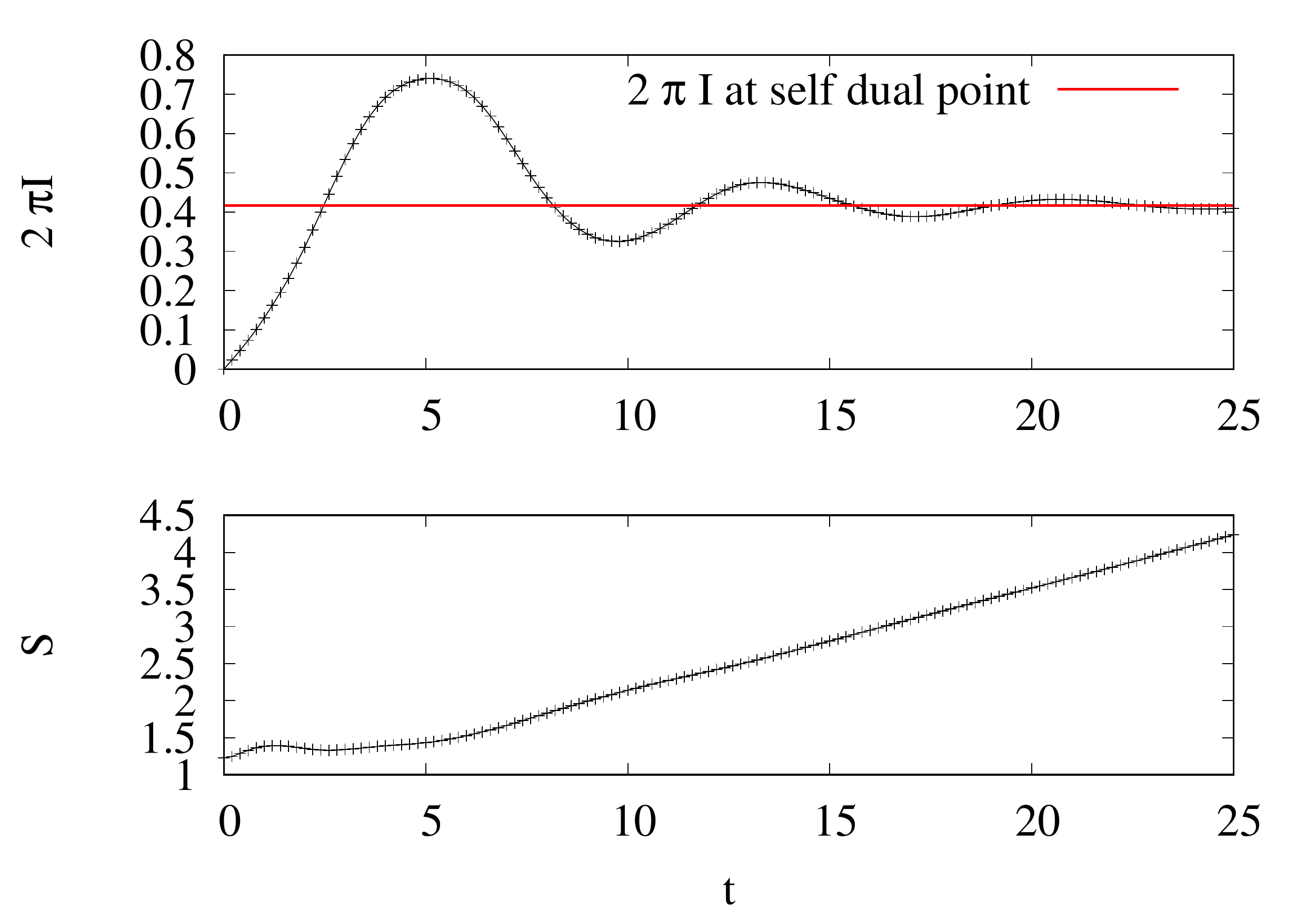}
\caption{Top panel: current $I(t)$. Bottom panel: entanglement entropy $S(t)$ between the left lead and the rest of the system.
The horizontal line in the top panel is the exact result at the self-dual point (see text), derived in Ref.~\cite{boulat_twofold_2008}.
Parameters: $U=2.0$, $J'=0.3$, $V=2.0$, $N=254$. }
\label{fig:C_E}
\end{figure}

The way the current in the center of the system approaches its steady value, after some damped oscillations, is shown in the upper panel of Fig.~\ref{fig:C_E}.

In many cases the oscillations which appear in the transient regime have a well-defined period $T_{\rm osc}$.
It was argued in Ref.~\cite{schneider_conductance_2006} that this period is simply given by the bias: $T_{\rm osc}=4\pi/V$.
Our data are in agreement with this result, from $U=0$ to large values of $U$.

\subsection{Entanglement entropy}
\label{ssec:entropy}

We denote by $S(t,R+\frac{1}{2})$ the von Neumann entanglement entropy of the set $A$ of sites located to the left of site $R$,
that is $A=\{-N/2, \dots, R\}$.
When no position is specified, $S(t)$ refers to the entanglement entropy $S(t,-\frac{1}{2})$ of the entire left (or right) lead.

The lower panel of Fig.~\ref{fig:profiles} illustrates how the entanglement profile evolves. The most striking feature is
the rapid growth of the entanglement entropy, which turns out to be linear in time for a given $r$ inside the ``light cone''
(see the lower panel in Fig.~\ref{fig:C_E}).
This linear growth of the entropy is well known in
the situations where some steady current is flowing through an impurity (or defect).
One can in particular mention the analogous case of a weak bond connecting two free leads, studied in detail in Ref.~\cite{eisler_entanglement_2012}.
In such situations, a  quantity of interest is the entropy {\em rate}, defined as $\alpha=\frac{d}{dt}S(t)$.

Since the computational cost of MPS-based methods grows exponentially with the amount of bipartite entanglement entropy in the system,
this linear entropy growth severely limits the longest times we can reach in the simulations.
This should be contrasted, for instance, with the slower logarithmic growth of the entropy in the case where 
the two leads are connected to the dot in the {\em absence of any bias}~\cite{kennes_universal_2014,vasseur_universal_2017}.
A logarithmic growth is in fact generic for local quenches in critical one dimensional systems \cite{calabrese_entanglement_2007,stephan_local_2011,eisler_entanglement_2012}.
The entropy growth is also logarithmic in the case of an XXZ spin chain ($|\Delta|<1$), with a translation invariant Hamiltonian, which is set in a current-carrying state using some domain-wall initial condition (equivalent to some non zero bias)~\cite{gobert_real-time_2005,sabetta_nonequilibrium_2013}.

The physical origin of the finite rate $\alpha$ is easy to understand in a noninteracting and semiclassical picture. Each incoming particle at energy $\epsilon$ has a finite probability $\mathcal{T}(\epsilon)$ to be transmitted to the other side of the dot, and a probability $\mathcal{R}(\epsilon)=1-\mathcal{T}(\epsilon)$ to be reflected (more details in Appendix~\ref{ssec:entropy_rate_ff}). After such a scattering event, the wave packet of the particle is split in two parts, one on each side of the dot, propagating in opposite directions.
In other words, the state of this particle is a quantum superposition of two terms, one in which the particle is in the left lead, and another one where the particle is in the right lead.
So, each such event contributes by an amount $\delta S=-\mathcal{T}(\epsilon)\ln \mathcal{T}(\epsilon) -\mathcal{R}(\epsilon)\ln \mathcal{R}(\epsilon)$ to the entanglement entropy between the two leads.
In presence of a finite steady current there is finite charge transmitted per unit of time, and hence a linear growth of the entropy $S(t)\sim \alpha t$ [except if $\mathcal{R}(\epsilon)=0$, as for $J'=J$]. In such a picture, the entanglement is directly related to quantum fluctuations of the transmitted charge, present as soon as $\mathcal{T}(\epsilon)$ is different from 1 and from 0,
and this is nothing but a manifestation of the relation between entanglement and charge fluctuations in free particle systems \cite{klich_quantum_2009,song_bipartite_2012,song_entanglement_2011}.
In contrast, if $U\ne0$, the entanglement growth is $\textit{a priori}$ not directly related to the partial transfer of the particles. 
We will indeed see in Sec.~\ref{ssec:entropy_rate} that one can be in a situation where $I$ goes to zero while the rate $\alpha$ stays finite.

The semiclassical description at $U=0$ can also be used to understand qualitatively the triangular shape of the entropy profile. Indeed,
in the limit where the bias $V$ is small compared to the bandwidth, all the relevant excitations propagate at the same group velocity ($\pm v_F=\pm 2J$). In that case, the degrees of freedom which contribute to the entanglement entropy form a ``train'' of left-moving wave packets with momenta centered around $- k_F$, and another train with right-moving wave packets centered at $+k_F$ \footnote{The average spacing $\lambda$ between the packets is proportional to the inverse of their momentum width: $\lambda=2\pi (\Delta k)^{-1}$ . The momentum width is related
to the energy width, $v_F \Delta k=\Delta\epsilon$, and this energy width is nothing but the bias, $\Delta\epsilon=V$.
We thus have $v_F \Delta k= V$. The number of incident particles per unit of time
is thus $v_F/\lambda=V/(2\pi)$, and this is consistent with the small $V$ limit of the current given in Eq.~\ref{eq:I_ff}.
Note that the actual distribution is in fact Poissonian.
}.
The important point is that each left-moving particle is entangled with one right-moving partner, located on the other side of the dot, at the same distance.
When performing a partition of the chain at a given time $t$ and at a given position $R$, the entanglement entropy that is predicted by the semiclassical description simply depends on the number
of entangled pairs which are separated by the partition. It is then straightforward to see that this leads to a {\em triangular} entropy profile,
with a spatial extension ranging from $R=-v_F t$ to $+v_F t$, and a height equal to $\alpha t$, with the rate $\alpha$ given in Eq.~(\ref{eq:alpha_ff_1}). Although this classical picture with noninteracting particles
does not apply to the interacting case, the numerical simulations show that the triangular shape of the entropy profiles is a robust feature, at least far enough from the dot.
The integrability of the IRLM implies that, in some sense, a particle picture is still applicable, even in presence of strong interactions. This property was crucial
to derive the results in Refs.~\cite{boulat_twofold_2008,fendley_exact_1995,fendley_exact_1995B}, and it might be related to the triangular profile observed here.
Closer to the dot, the profile is, however, not triangular, and this will be analyzed in Sec.~\ref{ssec:entropy_profile}.

It should finally be noted that this linear entropy growth is what makes this type of simulation difficult, since it forces the matrix dimensions in the MPS representation of the wave function to grow exponentially with time. Some more details on this point can be found in Appendix~\ref{ssec:mat_trunc}.

\section{Nonequilibrium steady states}
\label{sec:ness}

We discuss here the properties of the steady region which grows in the center of the system, and where local observables asymptotically become independent of time.
\subsection{Reminder on the scaling regime and the continuum limit of the IRLM}

As mentioned in the Introduction, we focus on the regime where
the free-fermion bandwidth $W=4J$ is larger than all the other energies in the problem, namely, $J'$, $V,$ and $U$ \footnote{In practice we take $J=1$ and restrict our numerical simulations to $V \lesssim 2$, $J'\lesssim 0.3$ and $U\leq 2.5$. Finite-$J'$ corrections start to be visible for $J'=0.5$. As for $U$, our data at $U=4$ are clearly outside the scaling regime.}.

In this regime, the microscopic details of the leads (like the band structure) are irrelevant, except for the Fermi velocity ($v_F=2J$), and their
gapless low-energy excitations are described in the continuum limit in terms of simple scale-invariant Hamiltonians for left- and right-moving relativistic fermions:
\begin{eqnarray}
 H_A^c&=&iv_F\int_{-\infty}^0 dr \left[\psi_L^\dagger(r)\partial_r\psi_L(r)-\psi_R^\dagger(r)\partial_r\psi_R(r)\right] \\
 H_B^c&=&iv_F\int_0^{\infty} dr \left[\psi_L^\dagger(r)\partial_r\psi_L(r)-\psi_R^\dagger(r)\partial_r\psi_R(r)\right].
\end{eqnarray}
$H_A^c$ describes the continuum limit of the left lead ($r<0$), $H_B^c$ describes that of the right lead ($r>0$), and $\psi_L$ and $\psi_R$
are the left- and right-moving fermion annihilation operators.
The coupling to the dot then takes the form
\begin{eqnarray}
H_d^c&=& -J_c'\left( \psi_L(0) + \psi_R(0)\right) d^\dagger +{\rm H.c.} \nonumber \\
&&+U_c\left(:\psi_L(0)^\dagger \psi_L(0):+:\psi_R(0)^\dagger \psi_R(0):\right)\nonumber\\
&&\times\left( d^\dagger d-\frac{1}{2}\right),
 \end{eqnarray}
where $d$ is the fermion operator associated to the dot.
The analysis of this model then usually proceeds by ``unfolding'' the two semi-infinite leads,
giving two infinite right-moving Fermi wires, but we will not pursue  this here.

Since the tunneling to the dot is a relevant interaction, 
a Kondo energy scale $T_B$ appears when the leads are connected to the dot, and most quantities are expected to follow
some single-parameter scaling
with $T_B$.
At energies that are small compared to the crossover scale $T_B$, the dot is hybridized with the leads, and the system effectively appears as a single chain (so-called ``healing''), whereas
the wires appear to be almost disconnected at energies much larger than $T_B$.
As for the interaction $U$, it corresponds to a marginal operator and therefore changes continuously the critical properties
of the model.
Although this is well established for equilibrium properties, it is less obvious
that $T_B$ also rules the out-of-equilibrium properties.
Such behavior is nevertheless verified for the current $I$, which, for a given $U$, takes the form $I/T_B=f(V/T_B)$ \cite{boulat_twofold_2008}.
The  role played by $T_B$ in the dynamics of the IRLM has also been investigated in the absence of any bias ($V=0$), in a local quench setup
when the leads are abruptly connected to the dot at $t=0$ \cite{kennes_universal_2014,vasseur_universal_2017}.

In Sec.~\ref{ssec:entropy} we show that the stationary rate $\alpha=\frac{d}{dt}S$ at which the entanglement entropy in the center grows with time obeys
a similar scaling form.

The energy $T_B$ is known to scale as $\sim J'^{\frac{1}{1-h}}$, where $h$ is the scaling dimension of the operator which, in the continuum description, describes the tunneling from the leads to the dot. This dimension $h$ depends on the interaction through $h=\frac{1}{4}+\left(\frac{1}{2}-\frac{U_c}{2\pi}\right)^2$, where $U_c$ is the interaction strength in the continuum limit~\cite{boulat_twofold_2008}.
Finding the precise way $U_c$ depends on the microscopic parameter $U$ of the lattice model would require to follow exactly
the renormalization-group flow going from the lattice model to the infra red fixed point, but there is no known method to do this exactly.
The analytical result of Ref.~\cite{boulat_twofold_2008} was obtained for the special value $U_c=\pi$, where
the model has some self-duality property and can be mapped onto the boundary sine-Gordon model.
Thanks to numerical simulations, it has been shown that the lattice model at $U= 2.0$ has a continuum limit which is close to the self dual point, where the exponent $h$ reaches a minimum~\cite{boulat_twofold_2008}.
Our data also confirm this result. We also improve quantitatively the connection between  interaction strength $U$ on the lattice, and its value $U_c$ in the infrared limit.

The exponent $h$ also appears in the limit of large (rescaled) bias $V/T_B$, where the steady current behaves as a power law:  $I\sim V^{-b}$ with $b=1-2h$ \cite{boulat_twofold_2008}.
This behavior will be checked numerically in detail in the next section (Sec.~\ref{ssec:sc}). For a given $U$, this offers a simple way  to extract $h$ and $b$ from the simulations, and then to define $T_B$ for each value of $J'$.

\subsection{Steady Current}
\label{ssec:sc}

\begin{figure}[h]
\includegraphics[height=0.5\textwidth, angle=270]{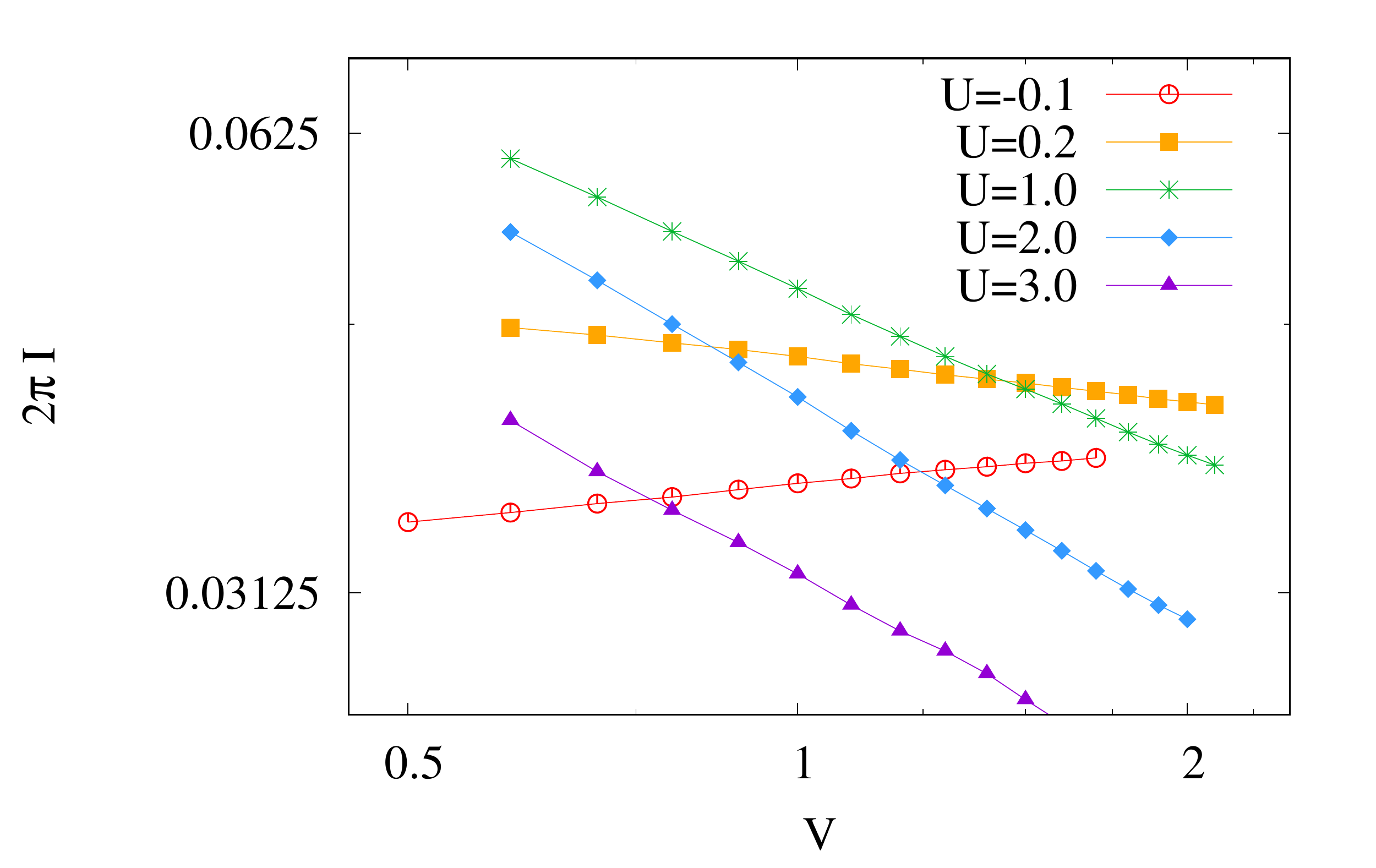}
\caption{
Stationary current $I$ as a function of $V$ for $J'=0.08$, for a few selected values of $U$.
A log-log scale is used to show the power law behavior of the current at large $V/T_B$ ($I\sim V^{-b}$), and to extract the associated exponent $b(U)$.
The values $b(U)$ extracted by these fits are displayed in Fig.~\ref{fig:exponent_b}.
}
\label{fig:current_log_scaling}
\end{figure}

\begin{figure}[h]
\includegraphics[height=0.5\textwidth, angle=270]{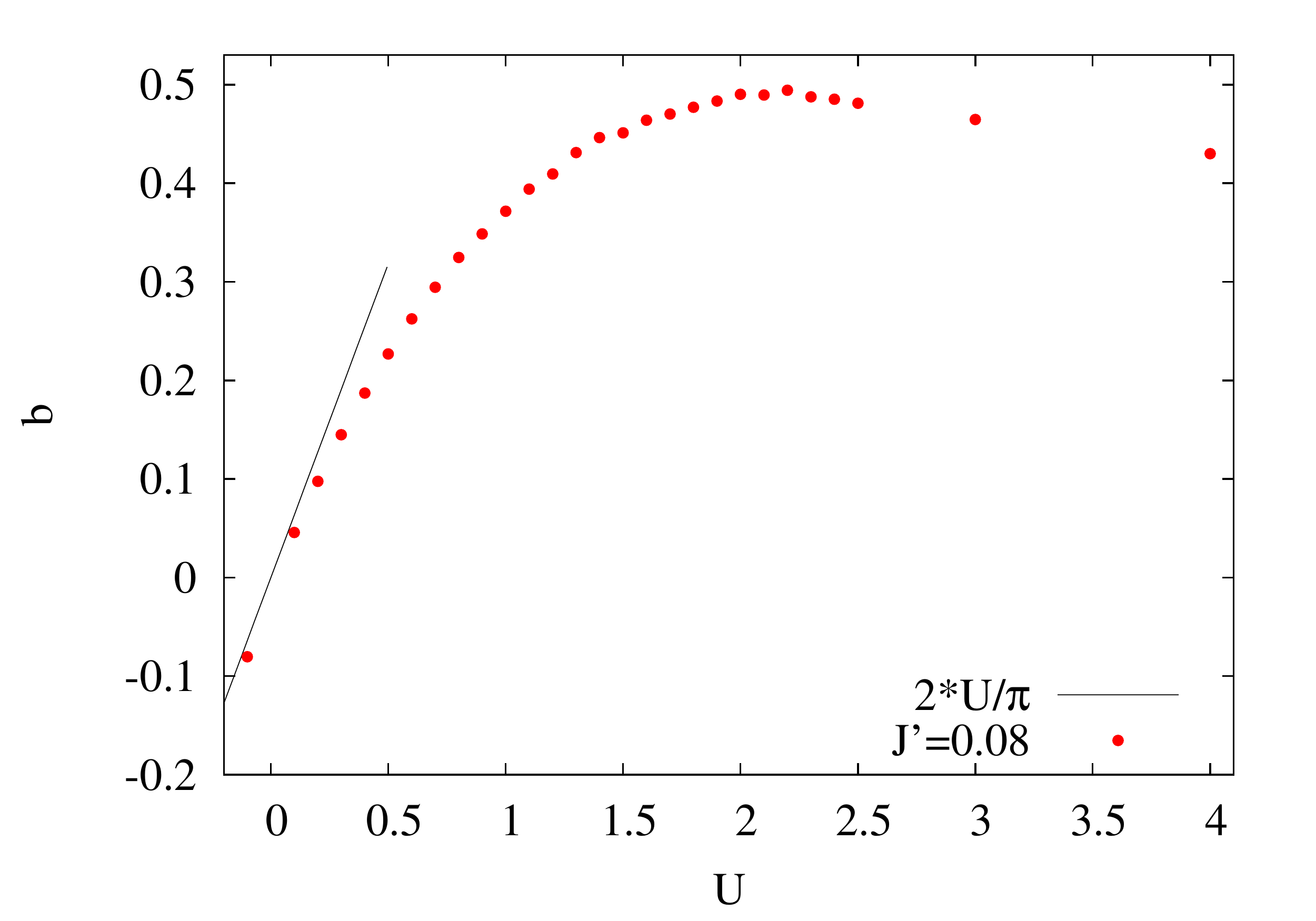}
\caption{
Exponent $b(U)$ as a function of the interaction strength $U$, obtained by fitting
the steady current $I$ to $I\sim V^{-b}$, at large $V/T_B$ (see Fig.~\ref{fig:current_log_scaling}). 
The full (black) line is $b=2U/\pi$, the result of some small-$U$ expansion~\cite{karrasch_functional_2010,vinkler-aviv_thermal_2014}.
}
\label{fig:exponent_b}
\end{figure}

Figure~\ref{fig:current_log_scaling} shows the  stationary current  as a function of the bias, for a few values of $U$ and $J'=0.08$.
A log-log scale is used to visualize the power-law behavior of the current $I\sim V^{-b(U)}$ at sufficiently large $V/T_B$, {\it i.e} small $J'$.
The  slope of the curve in log-log scale allows to determine the exponent $b(U)$ and $h(U)=\frac{1}{2}[1-b(U)]$. The results of these fits are shown in Fig.~\ref{fig:exponent_b}
\footnote{The value $J'=0.08$ appears to offer a good compromise to estimate $b$ in our simulations.
Indeed, we need a small $J'$ to be in the scaling regime, but  
the time to reach the steady regime (which increases when $J'$ decreases) should also  not become too large compared to $L/v_F$.}.
Then the scale $T_B$ is defined as $T_B(J',U)=(J')^\frac{1}{1-h(U)}=(J')^\frac{2}{1+b(U)}$ \footnote{We adopt the convention where the numerical prefactor in the definition of $T_B$ is set to 1.
}. From the analysis of the model in the continuum, we know that the exponent $b$ should reach a maximum value $b=\frac{1}{2}$ (equivalent to $h=\frac{1}{4}$) at the self-dual point.
The maximum value we obtain (Fig.~\ref{fig:exponent_b}) is $b=0.494$, which gives an estimate on our precision on this quantity.

\begin{figure}[h]
\includegraphics[height=0.5\textwidth, angle=270]{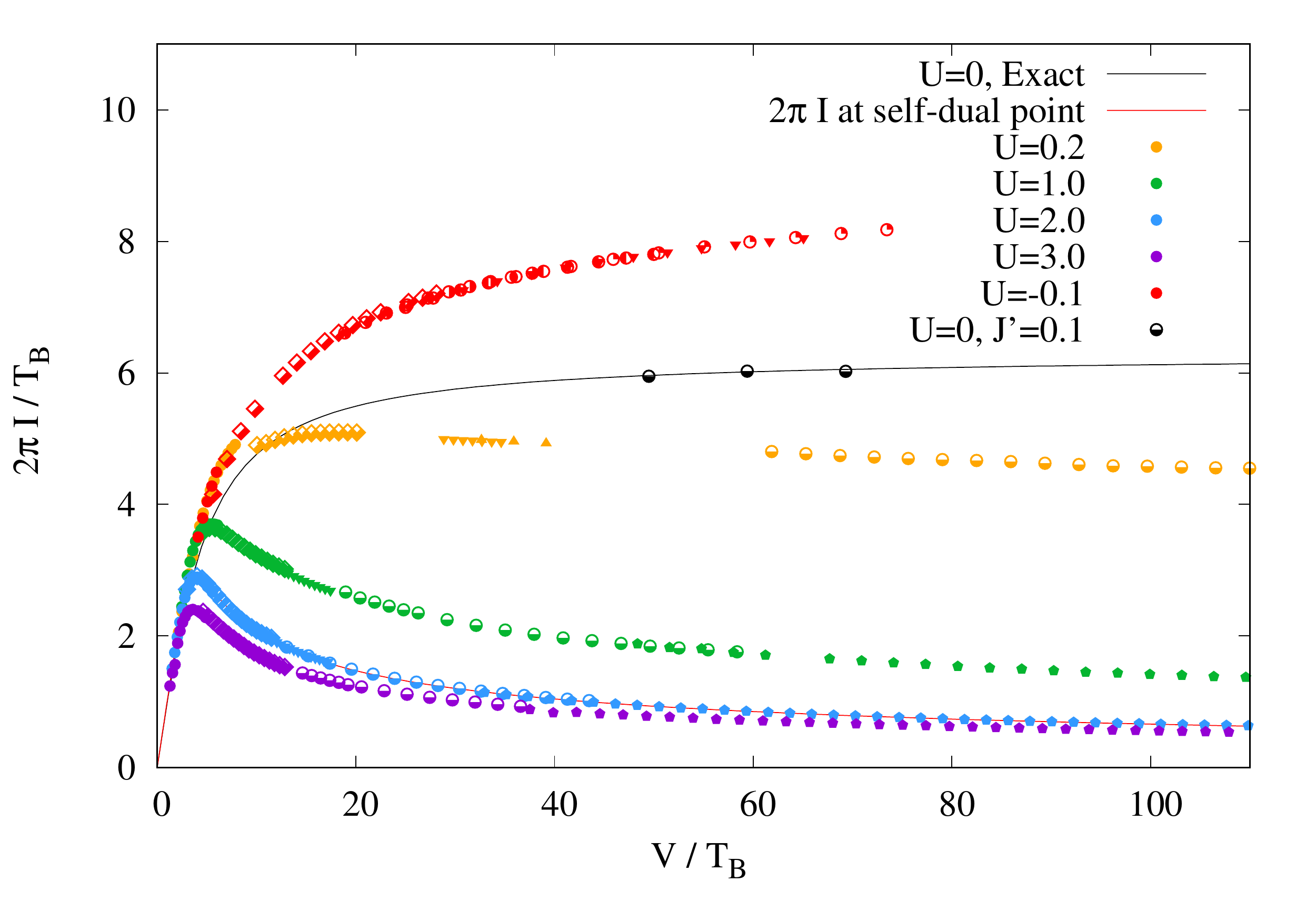}
\caption{Rescaled current $I/T_B$ as a function of the rescaled bias $V/T_B$, for different values of $U$.
The colors label the values of $U$, while the symbol shapes encode $J'$ (see the legend of Fig.~\ref{fig:current_scaling} for details).
For a given $U$, the results obtained for different values of $J'$ approximately collapse onto a single curve, as expected in the scaling regime.
The red line is the theoretical result for the self-dual point \cite{boulat_twofold_2008}.
The numerical data for $U=2.0$ appear to be in very good agreement with this theoretical curve.
The black line is the exact result for $U=0$ (Eq.~\ref{eq:I_ff}).
$T_B$ is defined as $T_B=(J')^\frac{1}{1-h(U)}$ where the exponent $h(U)$ is determined from the behavior of $I$ at large $V/T_B$ (see text
and Fig.~\ref{fig:exponent_b}).
}
\label{fig:current_scaling_all}
\end{figure}

\begin{figure}[h]
\includegraphics[height=0.5\textwidth, angle=270]{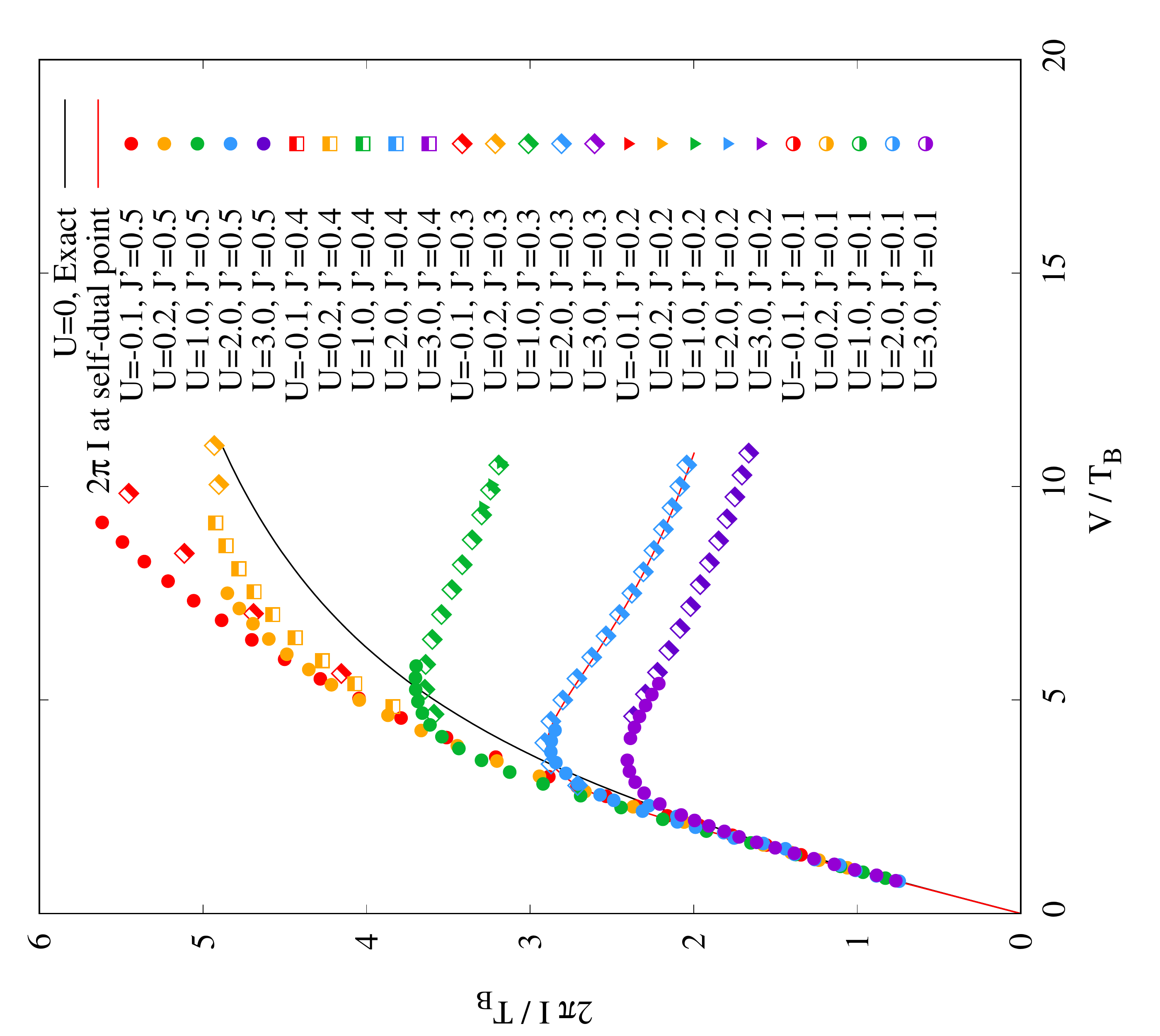}
\caption{Same as Fig.~\ref{fig:current_scaling_all} (rescaled current $I/T_B$ as a function of the  rescaled bias $V/T_B$), with a zoom on the low bias region.
The colors label the values of $U$, while the symbol shapes encode $J'$ (see the legend).
}
\label{fig:current_scaling}
\end{figure}

With the above $T_B$ one can define a rescaled current $I/T_B$ and rescaled bias $V/T_B$.
As shown in Figs.~\ref{fig:current_scaling_all} and \ref{fig:current_scaling},  we then observe a relatively good collapse, on a single master curve, of the data sets corresponding  to different $J'$ (for a given $U$).
This indicates that the lattice model is indeed close to the scaling regime, characterized by a single energy scale $T_B$.
For $U=2.0$ this has already been observed by Boulat {\it et al.} \cite{boulat_twofold_2008}, but  
thanks to longer simulations and larger systems, the present data
have some higher precision and we could extend the $I$-$V$ curves to larger values of $V/T_B$ (beyond 100) and for several values of $U$ from $-0.1$ to $3$.

The fact that the current  {\em decreases} with $V$ at large bias for $U>0$, called negative differential conductance, 
is a remarkable phenomenon due to the interaction (for $U\leq 0$ the current
is monotonically increasing), and has already been discussed
in \cite{boulat_twofold_2008}.

For small values of $U$ the exponent $b$ describing the current suppression at large bias has been computed using some
functional renormalization group method \cite{karrasch_functional_2010} or with some self-consistent conserving approximation~\cite{vinkler-aviv_thermal_2014}.
Using our convention for the lattice model, their result reads as $b= 2U/\pi +\mathcal{O}(U^2)$.
As shown in Fig.~\ref{fig:exponent_b}, this is consistent with our simulations.
Our results however appear to be slightly below this first order expansion in $U$. Together with the fact that the maximum of $b$
is found to be slightly below 0.5, this indicates that our procedure slightly underestimates the exponent.
This effect is presumably due to the fact that the calculations are performed using a finite $J'$ (0.08) and finite voltage $V$ (up to $\simeq2$).

\subsection{Entropy rate}
\label{ssec:alpha}

\begin{figure}[h]
\includegraphics[height=0.5\textwidth, angle=270]{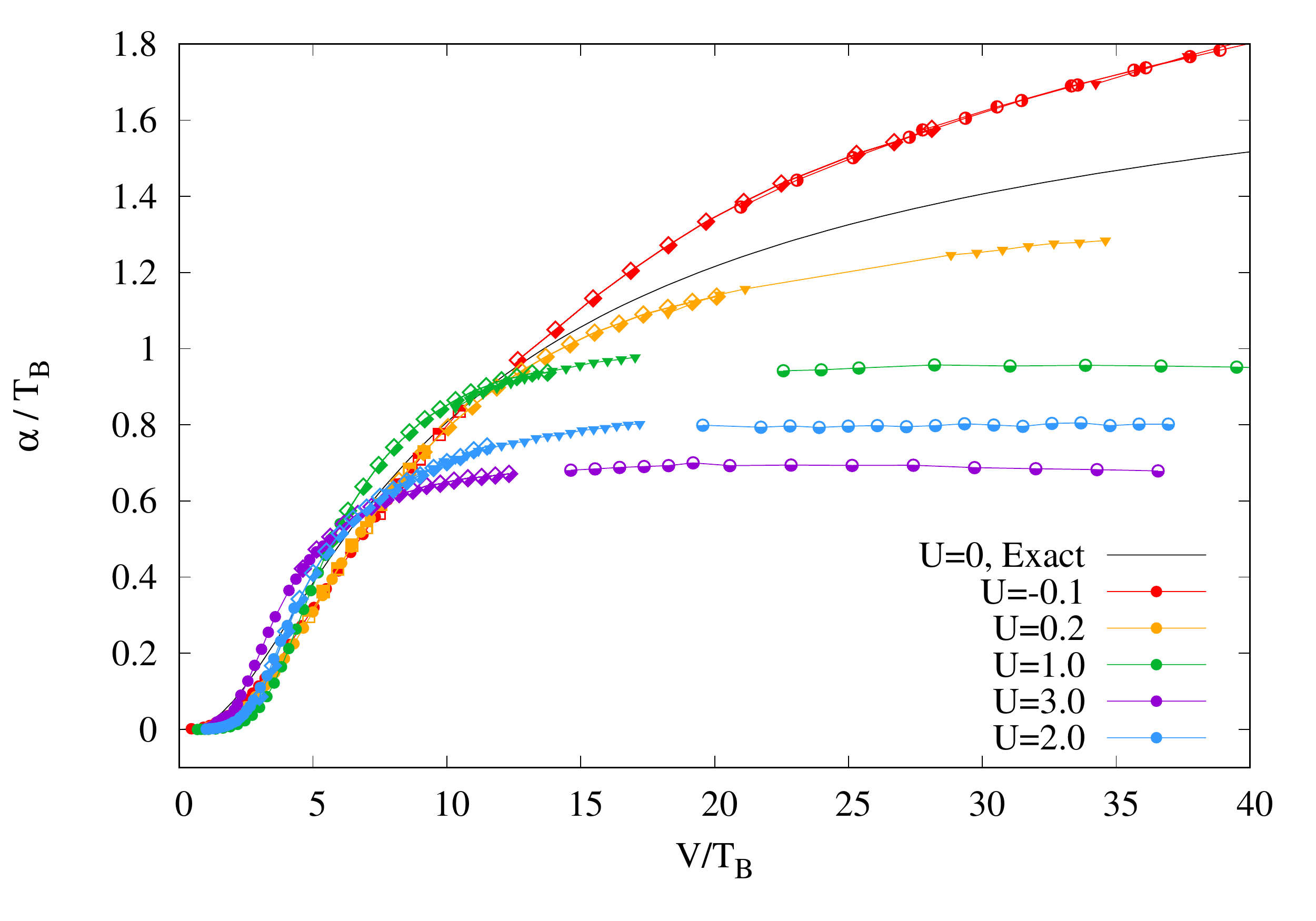}
\caption{Rescaled entropy rate, $\alpha/T_B$, as a function  of $V/T_B$, for several values of $U$.
The colors label the values of $U$, while the symbol shapes encode $J'$ (see the legend of Fig.~\ref{fig:current_scaling} for details).
The black line is the exact free-fermion result
(Eq.~\ref{eq:alpha_ff_2}).
}
\label{fig:entr_scal}
\end{figure}

\begin{figure}[h]
\includegraphics[height=0.5\textwidth, angle=270]{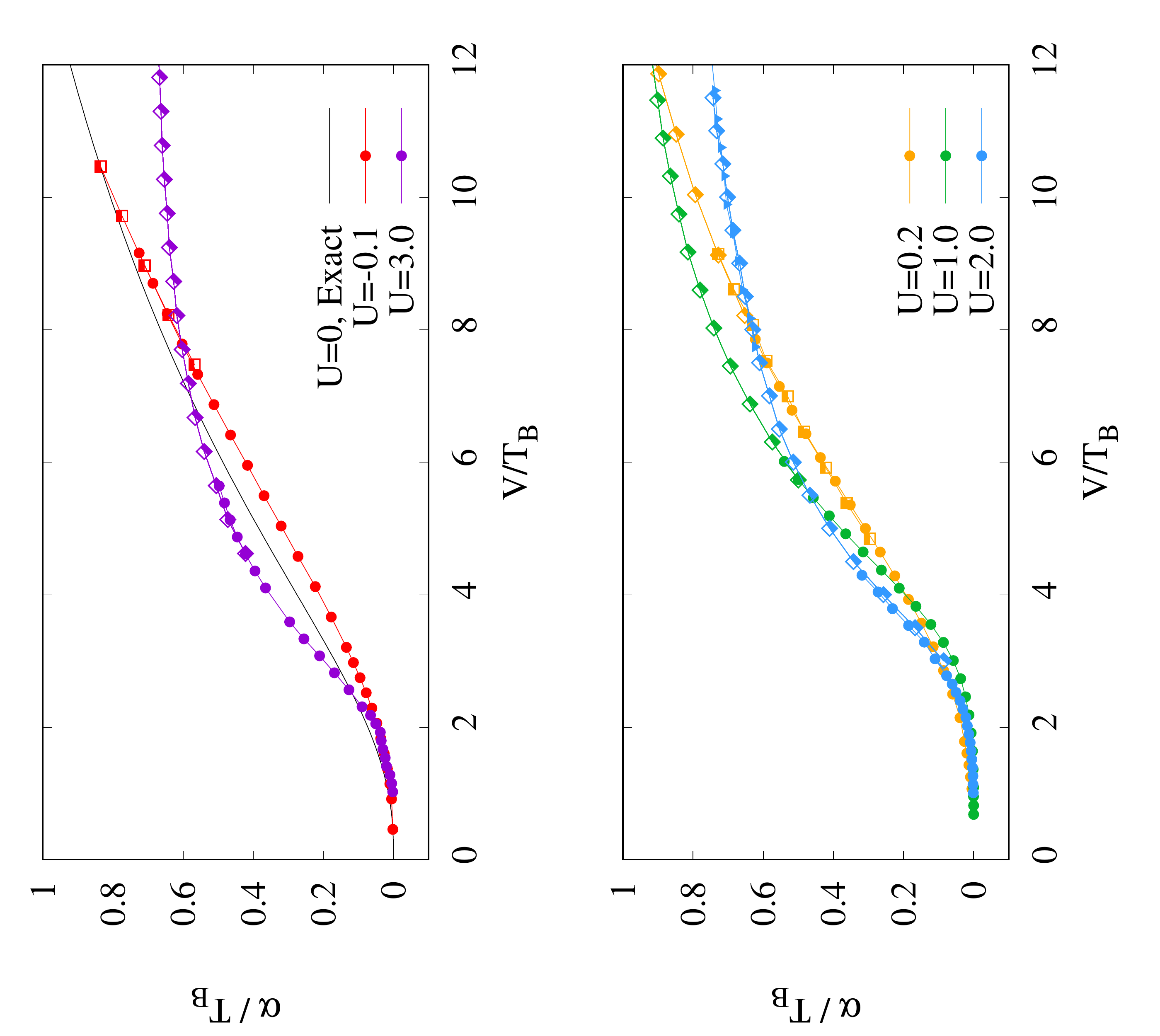}
\caption{Same as Fig.~\ref{fig:entr_scal}, with a zoom on the low bias region.
Top panel: $U=-0.1$, 0, and 3. Bottom panel: $U=0.2$, 1, and 2.
The colors encode the values of $U$, while the symbol shapes label the values of $J'$, as indicated in the legend
of Fig.~\ref{fig:current_scaling}. 
The black line (top panel) is the exact free-fermion result in the limit $J'\to0$
[Eq.~(\ref{eq:alpha_ff_2})].
}
\label{fig:entr_scal_2}
\end{figure}

We estimate the steady entropy  rate $\alpha$, defined in Sec.~\ref{ssec:entropy}, by fitting the long-time part of the entanglement entropy data $S(t)$
\footnote{
Like for the current $I(t)$, the entropy $S(t)$ often shows some oscillatory part,  $\sim \cos(Vt/2 + {\rm cst})$ (see Fig.~\ref{fig:delta_Jp=0.05}),
and we include such a term in the fitting function in order to improve the precision on $\alpha$.}.
As for the current, we consider the rescaled
entropy rate $\alpha/T_B$ as a function of the rescaled bias $V/T_B$. The results, plotted in Figs.~\ref{fig:entr_scal} and \ref{fig:entr_scal_2},
show that the data obtained for a given value $U$ but for different values of $J'$ collapse quite well onto a single master curve.
The values of $T_B$ used to construct the Figs.~\ref{fig:entr_scal} and \ref{fig:entr_scal_2} are the same as those used to analyze the scaling of the current.
From this point of view, the quality of the collapse for the entropy rate 
is quite remarkable since there is no adjustable parameter: the  scale $T_B$ was extracted  from the large-bias behavior of current only, and for a single value of $J'=0.08$. Note also that, to our knowledge, no exact result is known for the entropy rate when $U\ne 0$, even at the self-dual point.

The free-fermion result for the entropy rate (derived in Appendix~\ref{ssec:entropy_rate_ff})
converges to some finite constant, $\alpha/T_B\to 2$, at large rescaled bias. An important fact  we learn from the Fig.~\ref{fig:entr_scal}
is that,  in presence of interactions, $\alpha/T_B$ also saturates to some finite value at large $V/T_B$. 
This limiting value appears to be smaller than 2 when $U>0$. From our data the large bias behavior of the entropy rate when $U<0$ is not simple to guess.
It may diverge as $V/T_B\to \infty$, or it may saturate to some
value larger than 2.

For $U>0$, while the current decreases
to zero at large voltage (positive exponent $b$),  the entropy rate stays finite. This observation is somehow counter intuitive since
a vanishingly small amount of charge is transferred per unit of time from one wire to the other, while their entanglement entropy still grows at a finite rate.
In this large bias regime it is possible that the entanglement is generated by the density-density interaction $U$, without any actual transfer of particles from one lead to the other.
This question certainly deserves some further investigations.

\label{ssec:entropy_rate}

\label{ssec:entropy_rate_2}

\subsection{Stationary entropy profile}
\label{ssec:entropy_profile}

Several works have shown that, in Kondo-type problems, the entanglement entropy 
can be used to identify some spatial region of size $\xi \sim T_B^{-1}$ around the impurity (sometimes called Kondo "cloud").
See, for instance, Refs.~\cite{freton_infrared_2013,vasseur_universal_2017} concerning the IRLM, and Ref.~\cite{laflorencie_quantum_2016} for a more general review.
While these studies investigated the entanglement in the ground state of the model, here we instead look at the entropy profile
in some nonequilibrium steady state, in presence of a finite current.
 
\begin{figure}[h]
\includegraphics[height=0.5\textwidth, angle=270]{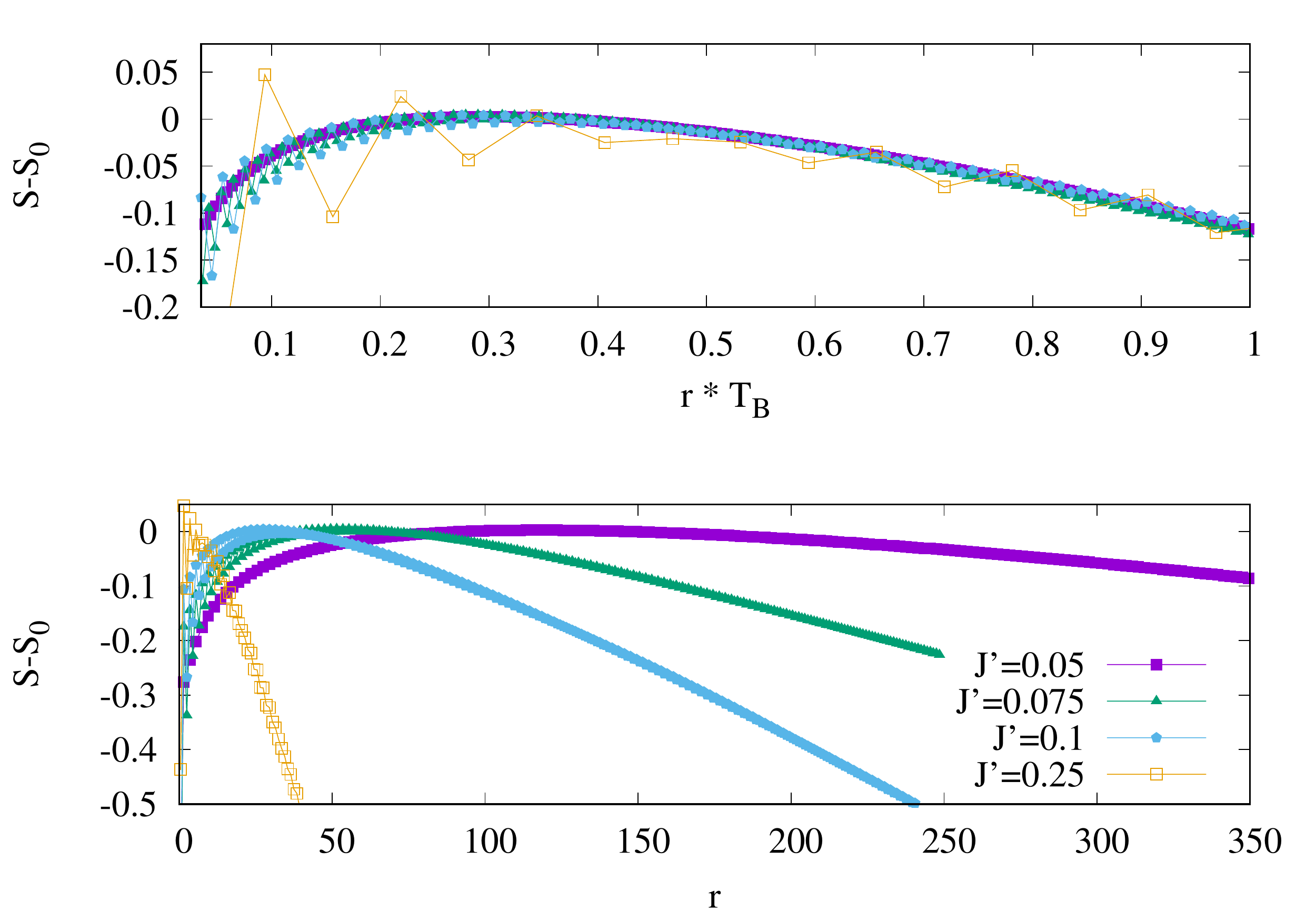}
\caption{Top: Entanglement entropy profiles for $U=0$, plotted as a function of the rescaled distance $r/\xi$, with $\xi=T_B ^{-1}$ and $T_B=J'^2$. 
Some constant value has been subtracted from each profile, to allow the data to (approximately) fall onto a single time-independent curve.
We only show here the right part of the profile $(r>0$), the other side ($r<0$) being symmetric. 
Bottom: same data plotted as a function of the bare distance $r$. System size is $N=6000$ and time is $t=1200$, 
values for $V(J')$: $V(0.25)=0.5, V(0.1)=0.08, V(0.075)=0.045, V(0.05)=0.02$ with ratio $ T_B ^{-1}(0.05) : T_B ^{-1} (0.075) : T_B ^{-1} (0.1) : T_B ^{-1} (0.25) = 25:4:2.25:1$.
In all cases the rescaled bias is $V/T_B$=8.0.
}
\label{fig:cloud_U=0}
\end{figure}

\begin{figure}[h]
\includegraphics[height=0.5\textwidth, angle=270]{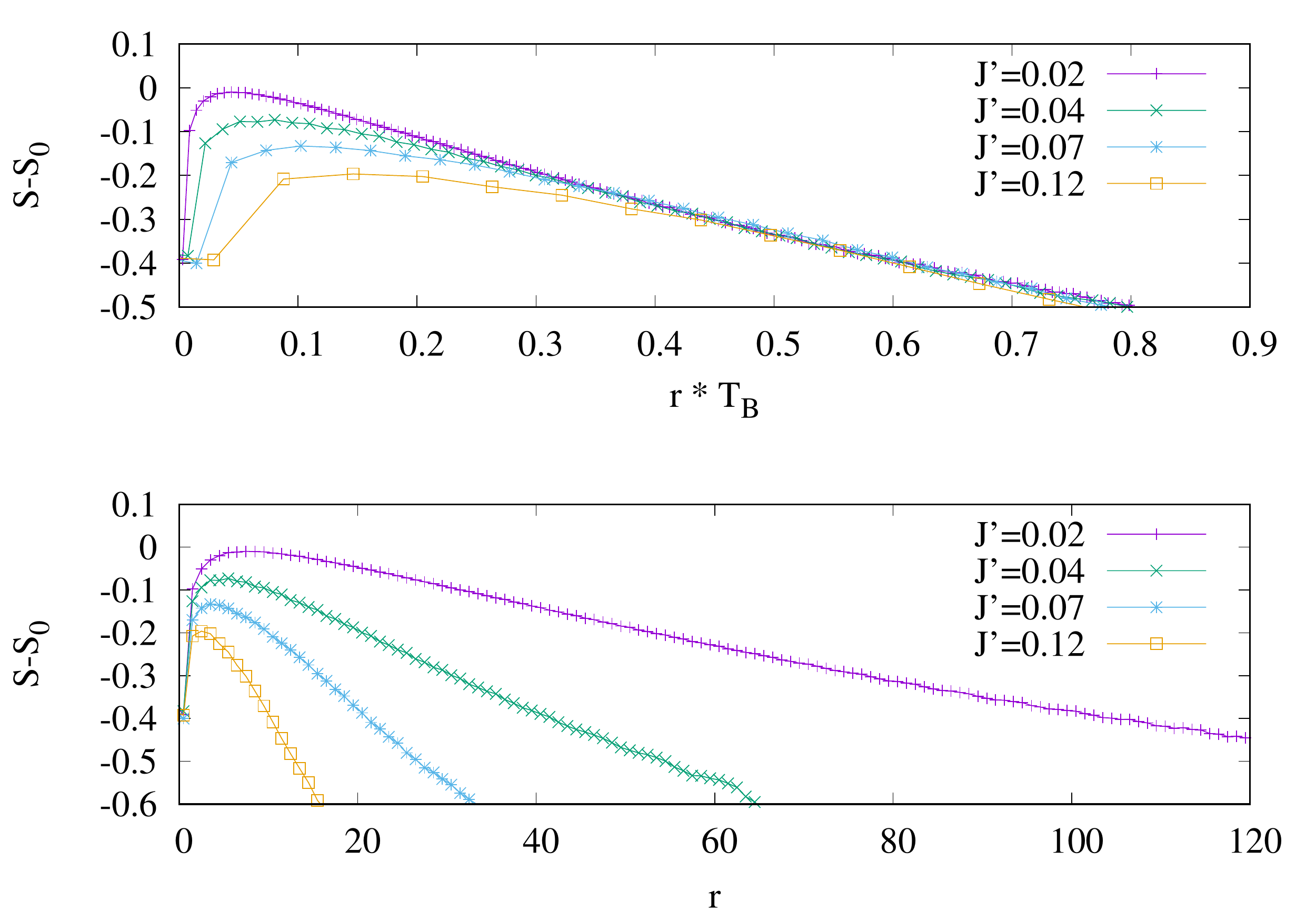}
\caption{
Same as Fig.~\ref{fig:cloud_U=0}, but for  $U=2.0$. System size is $N=402$ and $t=60$.
Values for $V(J')$: $V(0.12)=1.0, V(0.07)=0.5, V(0.04)=0.25, V(0.02)=0.1$ with ratio $ T_B ^{-1}(0.02) : T_B ^{-1} (0.04) : T_B ^{-1} (0.07) : T_B ^{-1} (1.2) = 10 : 4: 2 : 1 $. In all cases the rescaled bias is $V/T_B=17.1$. 
}
\label{fig:cloud_U=2}
\end{figure}

Our results are summarized in Figs.~\ref{fig:cloud_U=0} and  Figs.~\ref{fig:cloud_U=2}. As discussed in Sec.~\ref{ssec:entropy},
the global shape of the profile is approximately triangular, with some maximum value that grows as $\alpha t$ with a rate $\alpha \sim \mathcal{O}(T_B)$. In order to analyze 
more precisely the long time limit of the profiles, we subtract the value of the maximum of each profile.

When  plotted
as a function of the rescaled distance $r/\xi$, with $\xi=T_B ^{-1}$, the data for $U=0$ corresponding to different values of $J'$ (but same
$V/T_B$) collapse onto a single curve, at least approximately.
At distances from the dot which are large compared to $\xi$, the curve is  linear. In this region the collapse is a  consequence
of the fact that the entropy rate $\alpha$, and thus the slope of the profile, scales as $T_B$. However, the curve shows a nontrivial structure for distances from the dot that are of the order of $\xi$,
with some rounded maximum. We interpret this as some signature of the more complex correlations taking place in a nonequilibrium  Kondo 'cloud' of size $T_B^{-1}$.

The situation at $U=2$, displayed in Fig.~\ref{fig:cloud_U=2} is more intriguing. The use of the rescaled distance $r\cdot T_B$
still allows the data associated to different $J'$ (and thus different $T_B$) to collapse on a straight line for distances that are sufficiently large.
But since the entanglement entropy
is just a linear function of the distance to the dot when $r$ is sufficiently large, this collapse
only reflects the fact that the slope of the entropy profile is proportional
to $T_B$ (which we know already, since the entropy rate scales as $T_B$ and the Fermi velocity is equal to 2). Closer to the dot, there is however no clear convergence for \mbox{$r\cdot T_B\lesssim 0.4$}. More precisely, the distance $r_{\rm max}$ at which the entropy profiles reaches a maximum clearly grows when $J'$ goes to zero, but at some rate which is slower than $T_B ^{-1}$. So, from the present data at $U=2$ (and $V/T_B=17.1$), there is no clear evidence of some Kondo-like spatial structure of size $\sim T_B ^{-1}$ in the stationary entropy profiles, as observed in the free case. It could however be that some longer times are required to achieve some profile collapse in this regime. Some further systematic investigations, as a function of $U$ and $V$ and $J'$, would be required to elucidate this point.

\section*{Conclusions}

We have  analyzed a  number of numerical results
concerning the steady state properties of the IRLM in the scaling regime: $I$-$V$ curves, exponent $b(U)$ and the Kondo scale
$T_B$, and entropy rate $\alpha$. We could in particular obtain accurately the rescaled 
$I$-$V$ curves and the entropy rate $\alpha$ in some large range of the rescaled bias (up to $V/T_B=100$), and for several values of the interaction strength $U$.

We hope that this study will trigger some further investigations  using analytical techniques, since our results might be compared
quantitatively to field-theory results obtained  thanks to the integrability of the model in the continuum limit, along the lines of what has already been done at the self-dual point~\cite{boulat_twofold_2008},
or for the boundary sine-Gordon model~\cite{fendley_exact_1995,fendley_exact_1995B}. It would also be very interesting to elucidate
qualitatively the mechanisms at work in the limit of large bias and $U>0$, to understand how a finite entropy rate can coexist with a vanishingly small current. Finally, our
data can be used to benchmark new approximations for quantum impurity models, or when developing numerical schemes which can deal
with long time evolutions in interacting problems with linear entanglement entropy growth.

\section*{Acknowledgements}

We are indebted to H. Saleur for valuable suggestions and discussions, as well as for sharing some of his unpublished notes.
KB is grateful to Soltan Bidzhiev for encouragements and support in this project.

\appendix

\section{Details about the numerical simulations}
\label{sec:numerics}

\subsection{Initial state and Unitary evolution}

The initial state is computed using a conventional DMRG procedure, and the sweeps are stopped when the energy variation becomes smaller than $\Delta E = 10^{-10}$. During this initial state calculation the level of MPS truncation
is the same as during the time evolution, and is determined by some maximum value $\delta$ for the discarded weight, typically set to $10^{-7}$.

Most of the data are calculated for chains of length between $N=120$ and 200. Unless mentioned otherwise, the time evolution is performed with time steps  $\tau = 0.2$  and we used the MPO-based
approximation to $\exp(-iH\tau)$ that is noted $W^{\rm II}$ in Ref.~\cite{zaletel_time-evolving_2015}.
As explained in this previous work, one can combine several $W^{\rm II}$ to get a smaller
error:
\begin{equation}
 W^{\rm II}(\tau_1)W^{\rm II}(\tau_2)\cdots W^{\rm II}(\tau_n)=\exp(H \tau)+\mathcal{O}(\tau^p)
 \label{eq:trotter}
\end{equation}

\begin{figure}[h]
\includegraphics[height=0.5\textwidth, angle=270]{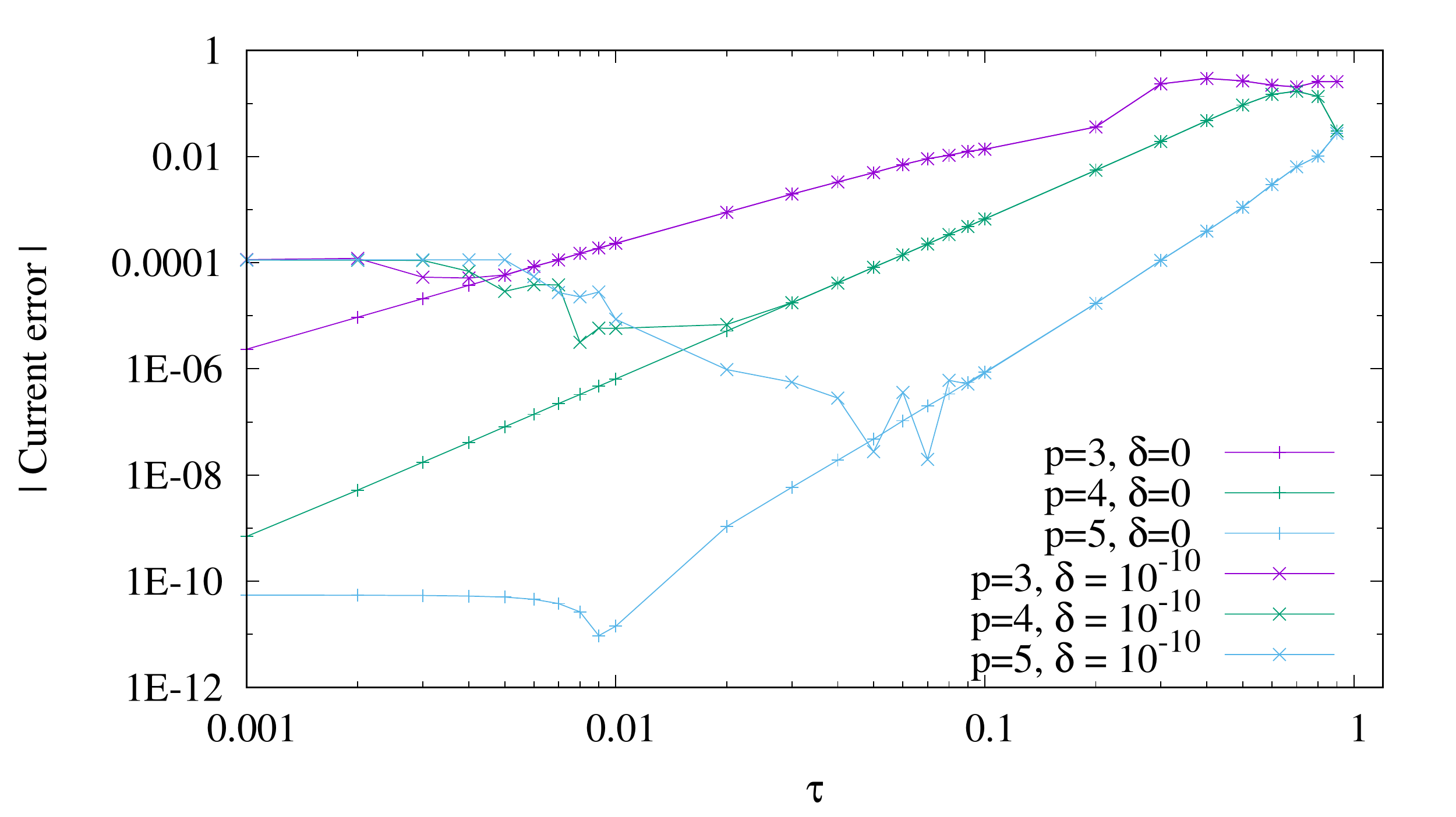}
\caption{Absolute value of the error for the expectation value of the current flowing through the dot, for $ N=6, U=0, V=1.0, t=100, J=0.5, J'=0.5$ 
The error is computed by comparing the MPS simulations with an exact free-fermion calculation.
For small $\tau$ the main source of error is no longer the finite $\tau$, but the successive truncations in the Schmidt decompositions (SVD).
This can be seen by comparing the data with cut-off parameter $\delta=10^{-10}$ to those without any truncation ($\delta=0$).
When the error becomes extremely small, of the order of $10^{-10}$, it stops decreasing with $\tau$.
This is due to the finite floating point precision of the machine.
}
\label{fig:trotter}
\end{figure}

The simplest case is to use $n=2$ steps to reduce the error to $\mathcal{O}(\tau^3)$, and
one solution is:
\begin{eqnarray}
\tau_1&=&\frac{1+i}{2}\tau \nonumber\\
\tau_2&=&\frac{1-i}{2}\tau.
\end{eqnarray}
We have computed two other solutions, corresponding to $p=4$ and $p=5$. These were mentioned in Ref.~\cite{zaletel_time-evolving_2015}, but not given explicitly.
The first one requires $n=4$ and is given by
\begin{eqnarray}
\tau_1&=&\frac{1}{4}\left( -\frac{1+i}{\sqrt{3}}+1-i \right)\tau \nonumber\\
\tau_2&=&i \tau_1   \nonumber\\
\tau_3&=&-i \bar \tau_1   \nonumber\\
\tau_4&=&  \bar \tau_1.
\end{eqnarray}
The next one, with an error scaling as $\mathcal{O}(\tau^{p=5})$ requires $n=7$ steps. It was obtained
numerically using $\textit{Mathematica}$:
\begin{eqnarray}
\tau_1/\tau = 0.2588533986109182	+0.0447561340111419i, \nonumber\\
\tau_2/\tau =-0.0315468581488038	+0.2491190542755632i, \nonumber\\
\tau_3/\tau = 0.1908290521106672	-0.2318537492321061i, \nonumber\\
\tau_4/\tau = 0.1637288148544367,	\nonumber\\
\tau_5=\bar{\tau_3}, \nonumber\\
\tau_6=\bar{\tau_2}, \nonumber\\
\tau_7=\bar{\tau_1} \nonumber. \\
\label{eq:trotter4}
\end{eqnarray}
It should be noted that these solutions are not unique, but we have selected the ones where the error terms, of the order $\mathcal{O}(\tau^p)$, have the smallest prefactors.
We have checked the precision of these three different Trotter schemes on a small free-fermion system with 6 spins, and compared the results for the current
at time $t=100$ against the exact free-fermion solution. The results are shown
in Fig.~\ref{fig:trotter}. As expected, the total error scales as $\tau^{p-1}$, due to the fact that
the total number of steps is $t/\tau$ and each step contributes an amount of order $\mathcal{O}(\tau^p)$
to the error.

\subsection{Matrix truncations}
\label{ssec:mat_trunc}

\begin{figure}[h]
\includegraphics[height=0.5\textwidth, angle=270]{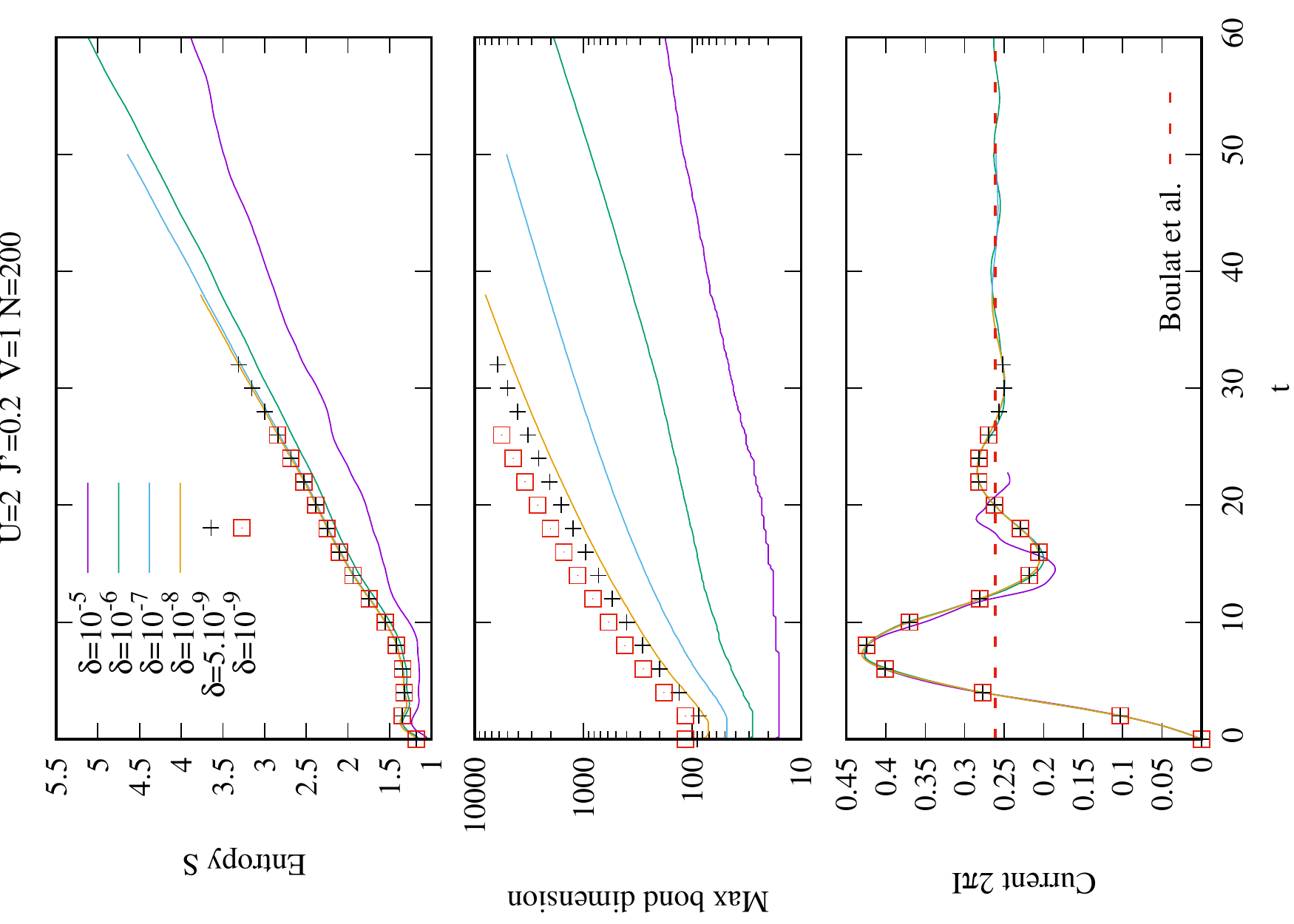}
\caption{Top panel: entanglement entropy of the left lead as a function of time for different values of the
SVD truncation parameter $\delta$ which defines the maximum discarded weight at each time step.
Middle: maximum MPS bond dimension.
Bottom: current flowing through the dot. Parameters of the model: $N=200$, $U=2$, $J'=0.2$ and $V=1$.
The (red) horizontal dotted line indicates the exact value of the steady current at the self-dual point \cite{boulat_twofold_2008}.
For these parameters, the upper panel shows that a truncation value as low as $\delta=10^{-7}$ is required to obtain some
accurate value for the entanglement entropy, leading to large bond dimensions (above 1000). In contrast,  $\delta=10^{-6}$ appears to be enough to
estimate the current correctly.
Simulations performed with Trotter step $\tau=0.2$.
}
\label{fig:delta_Jp=0.2}
\end{figure}

\begin{figure}[h]
\includegraphics[height=0.5\textwidth, angle=270]{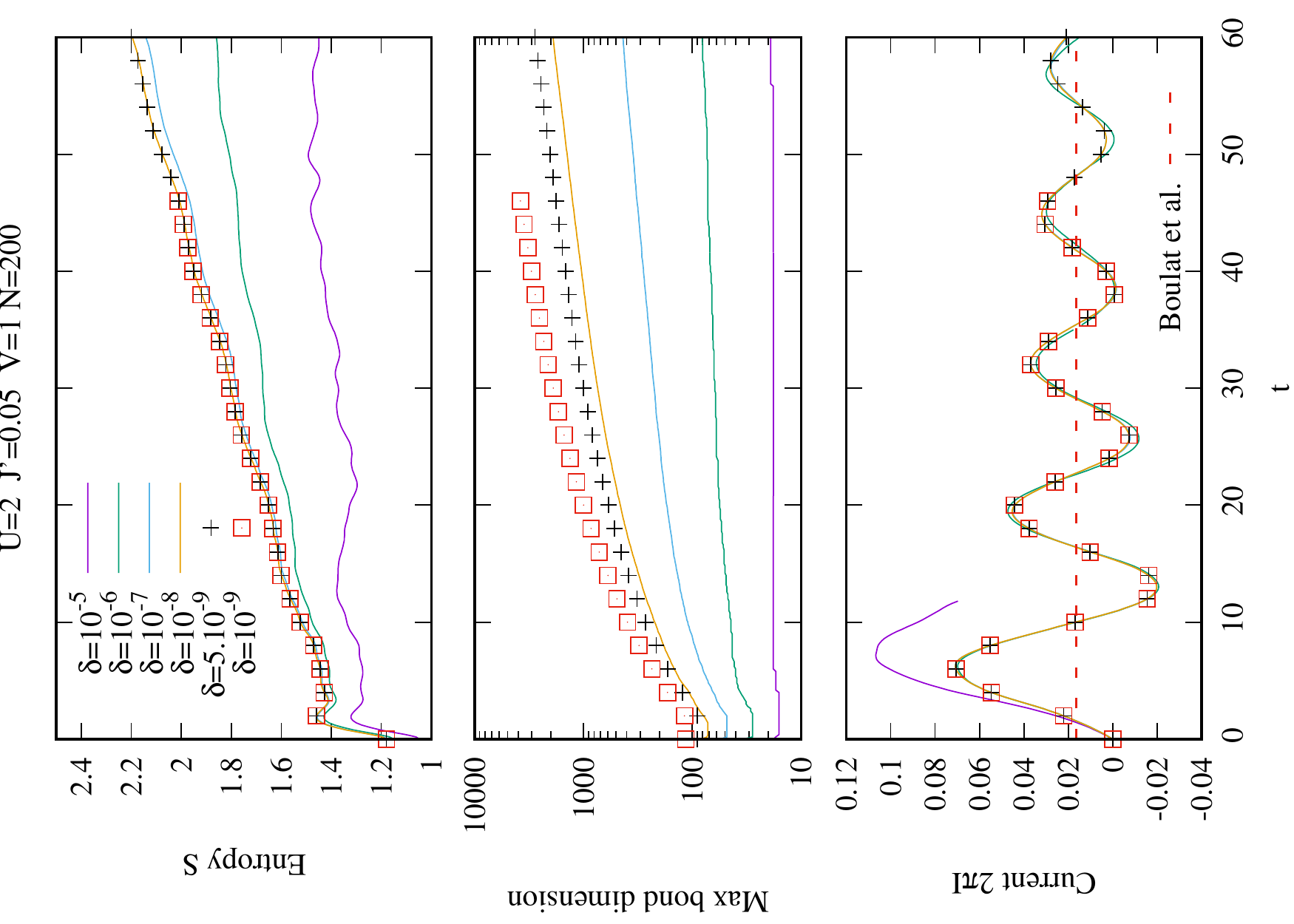}
\caption{Same as Fig.~\ref{fig:delta_Jp=0.2}, for $J'=0.05$.
For these parameters, a truncation value as low as $\delta=10^{-8}$ is required to obtain some
accurate value for the entanglement entropy, up to $t=60$.
One observes some small amplitude oscillations in $S(t)$, with the same period $T_{\rm osc}=4\pi/V$, as for $I(t)$.
}
\label{fig:delta_Jp=0.05}
\end{figure}

During the initial DMRG sweeps (initial state calculation) and during the time evolution, the bond dimensions in the MPS are controlled using a cut off equal to $\delta$ on the discarded weight (the
sum of the discarded Schmidt values should be equal to or smaller than $\delta$).
The effect of the truncation parameter $\delta$ is illustrated
in Figs.~\ref{fig:delta_Jp=0.2} and \ref{fig:delta_Jp=0.05}.
In this case a value $\delta=10^{-6}$ would provide a sufficient precision on the current (lower panel),
but it should be noted that obtaining a precise value for the entropy rate (upper panel of Fig.~\ref{fig:delta_Jp=0.2})
requires working with a smaller $\delta$.
For smaller $J'$, as shown in Fig.~\ref{fig:delta_Jp=0.05}, one has to use $\delta=10^{-8}$ in order to get some accurate estimate of the entropy rate.
In that case the bond dimension can exceed 1000 at time $\simeq 40$.

\section{Steady state in the free-fermion case}
\label{sec:ff}
\subsection{Steady current}
\begin{figure}[h]
\includegraphics[height=0.5\textwidth, angle=270]{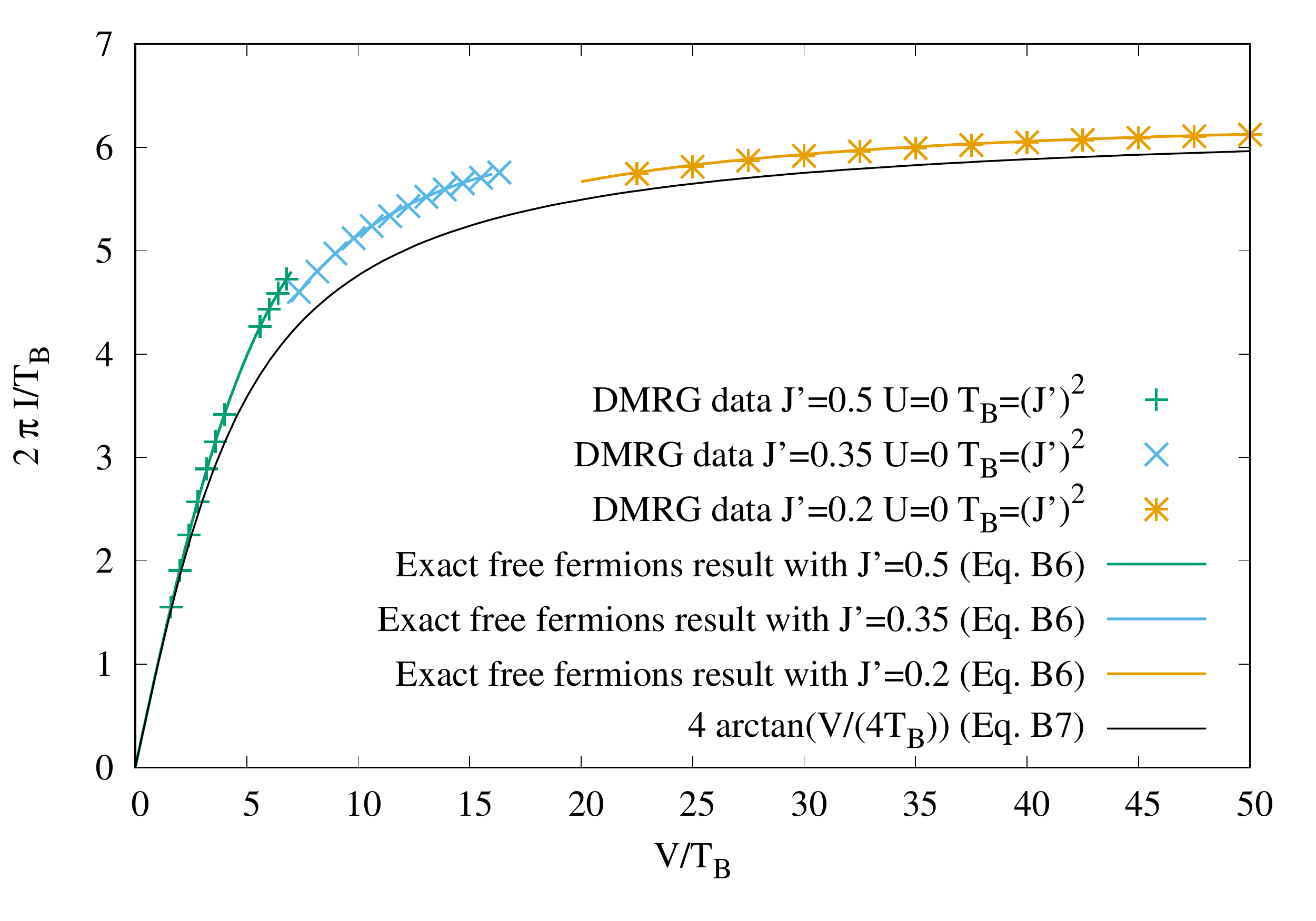}
\caption{Rescaled steady current as a function of the rescaled voltage in the free case ($U=0$), using $T_B=J'^2$.
The symbols represent the DMRG data for three different values of $J'$, and the full lines
are respectively the exact result (Eq.~\ref{eq:I_FF_exact}) for $J'=0.5$ (green), for $J'=0.35$ (blue), for $J'=0.2$ (orange), and the limit when $J'$ is small (Eq.~\ref{eq:I_ff}, black).
The DMRG data are in perfect agreement with the exact results, but since chosen $J'$ are not very small one observes some deviations
from Eq.~\ref{eq:I_ff}.
}
\label{fig:I_U=0}
\end{figure}

In the free-fermion case ($U=0$), the transmission and reflexion coefficients $\mathcal{T}(k)$ and $\mathcal{R}(k)$
for an incident fermion with momentum $k$ 
have been computed by Branschadel {\it et al.} \cite{branschadel_numerical_2010}:
\begin{eqnarray}
\mathcal{T}(\epsilon)&=&\frac{1-\left(\epsilon/(2J)\right)^2}
	  {1+\epsilon^2(J^2-2J'^2)/\left(4J'^4\right)} \label{eq:T} \\
	  \mathcal{R}(\epsilon)&=&1-\mathcal{T}(\epsilon)  \\
\epsilon(k)&=&-2J\cos(k)
 \end{eqnarray}
Combined with a Landauer approach~\cite{landauer_spatial_1957,imry_conductance_1999,nazarov_quantum_2009} this gives the steady current:
\begin{equation}
 I=\int_{k_F^-}^{k_F^+} \frac{dk}{2\pi} \mathcal{T}(\epsilon(k)) v(k)
\end{equation}
where the Fermi momenta in both leads are related to the voltage bias through $\epsilon(k_F^{\pm})=\pm V/2$
and the group velocity $v(k)$ is, by definition, $v(k)=\frac{\partial \epsilon(k)}{\partial k}$.
Changing the integration variable from $k$ to $\epsilon$ gives the standard result :
\begin{equation}
 I=\int_{-V/2}^{V/2} \frac{d\epsilon}{2\pi} \mathcal{T}(\epsilon). \label{eq:IVFF}
\end{equation}
For the IRLM, using Eq.~\ref{eq:T}, the integral gives:
\begin{eqnarray}
 2\pi I&=&-V\frac{J'^4}{a}\nonumber \\
 &&+4J'^2\frac{\left(1-J'^2\right)^2}{a^\frac{3}{2}}\arctan\left(\frac{\sqrt{a}}{4J'^2}V\right) \\
  a&=&1-2J'^2 \nonumber
 \label{eq:I_FF_exact}
\end{eqnarray}
where we assumed $J=1$, $V<4$ and $J'^2<1$. This function is plotted in Fig.~\ref{fig:I_U=0} for three different values of $J'$.
In the (scaling) limit where $J'\ll 1$ we define $T_B=(J')^2$
and the Eq.~\ref{eq:I_FF_exact} simplifies to
\begin{equation}
 \frac{2\pi I}{T_B}  = 4  \arctan\left(\frac{V}{4T_B}\right),
 \label{eq:I_ff}
\end{equation}
in agreement with Ref.~\cite{boulat_twofold_2008}.
Note that the expression $I(t)$ of the current at finite time and $U=0$ can be found in the appendix of Ref.~\cite{vinkler-aviv_thermal_2014}.
It shows damped oscillations at frequency $T_{\rm osc}=4\pi/V$, a relaxation time $\mathcal{O}(J'^2)$, and converges
to Eq.~\ref{eq:I_ff} when $t\to\infty$.

\subsection{Density drop across the dot}
\label{ssec:dens_drop}

We make the approximation that  the fermions are point-like particles which propagate ballistically in leads, at some  group velocity $v(k)$.
This is a semi-classical approximation (called hydrodynamical approximation in Ref.~\cite{antal_logarithmic_2008}) where each particle has a well defined position {\it and} momentum.
In this approximation, each lead becomes homogeneous in the steady regime and the system is  described by some occupation numbers $n(k)^{R/L}$ in both leads.
Taking into account the initial momentum distributions and the scattering on the dot, we obtain:
\begin{eqnarray}
 n(k)^L&=&\left\{
\begin{array}{cl}
0, & {\rm if}\;\; k^+_F < k \\
1, & {\rm if}\;\; -k_F^-<k<k^+_F \\
\mathcal{R}(k), & {\rm if}\;\; -k_F^+<k<-k^-_F \\
0, & {\rm if}\;\; k<-k^+_F \\
\end{array}\right. \\
 n(k)^R&=&\left\{
\begin{array}{cl}
0, & {\rm if}\;\;  k^+_F < k \\
\mathcal{T}(k), & {\rm if}\;\; k_F^-<k<k^+_F \\
1, & {\rm if}\;\; -k_F^-<k<k^-_F \\
0, & {\rm if}\;\; k<-k^-_F. \\
\end{array}\right.
\end{eqnarray}
The total density in each lead is then obtained by integrating the distributions
 above. Using the fact that $k_F^+ + k_F^-=\pi$ and $\mathcal{R}(k)+\mathcal{T}(k)=1$ we find:
\begin{eqnarray}
 \rho^L&=&\frac{1}{2}+\int_{k_F^-}^{k_F^+} \frac{dk}{2\pi} \mathcal{R}(\epsilon(k)) \\
 \rho^R&=&\frac{1}{2}-\int_{k_F^-}^{k_F^+} \frac{dk}{2\pi} \mathcal{R}(\epsilon(k)).
\end{eqnarray}
These densities describe the parts of the leads that are sufficiently far from the
dot ($|r|\gg 1$), and at times that are sufficiently long ($t\gg r$), so that the growing quasi-steady region has reached
$r$ (and $-r$). On a finite system we should additionally require $2Jt\lesssim N/2$ ($2J$ being the fastest group velocity).

We thus expect some density drop
\begin{equation}
\rho^L-\rho^R=2\int_{k_F^-}^{k_F^+} \frac{dk}{2\pi}\mathcal{R}(\epsilon(k))
\end{equation}
across the dot.

\subsection{Entropy rate}\label{ssec:entropy_rate_ff}

In analogy with the Landauer approach for the current,
the stationary entropy rate $\alpha$ can be obtained analytically.
Since each incident particle at energy $\epsilon$ contributes to the entropy
by an amount $\delta S=-\mathcal{T}(\epsilon)\ln \mathcal{T}(\epsilon) -\mathcal{R}(\epsilon)\ln \mathcal{R}(\epsilon)$
(its wave packet is split into a reflected part and a transmitted part), we get:
\begin{equation}
 \alpha = - \frac{1}{2\pi}\int_{-V/2}^{V/2} d\epsilon\left[
  \mathcal{T}(\epsilon) \ln \mathcal{T}(\epsilon)
  +
  \mathcal{R}(\epsilon) \ln \mathcal{R}(\epsilon)
 \right].
 \label{eq:alpha_ff_1}
\end{equation}
The result above has been checked numerically against numerical solution of dynamics for the free-fermion problem.

In the scaling regime $J'$ is small, the energy $\epsilon$ are small (because the bias $V$ is small), but the ratio $\epsilon/J'^2$ is of order one.
In this limit the transmission coefficient (Eq.~\ref{eq:T}) becomes:
\begin{equation}
 \mathcal{T}(x)=\frac{1}{1+x^2}
\end{equation}
where $x=\epsilon/(2T_B)$ and $T_B=J'^2$. With a change of variable the entropy rate (Eq.~\ref{eq:alpha_ff_1}) can be expressed as:
\begin{eqnarray}
 \alpha(v)&=&\frac{2T_B}{2\pi} \int_{-v/4}^{v/4} dx\left(
 \frac{1}{1+x^2}\ln (1+x^2) \right.\nonumber \\
 &&
  \left.
  + \frac{x^2}{1+x^2}\ln\left(\frac{1+x^2}{x^2}\right)  	
 \right),
\end{eqnarray}
where $v=V/T_B$.
The integral can be computed explicitly and the final result is
\begin{eqnarray}
 \frac{\alpha(v)}{T_B}&=&\frac{2}{\pi}\left[2\left( 1 +\ln(v/4) \right) 
\arctan \left( v/4 \right) \right.\nonumber \\
&&+\frac{1}{4}\,v\ln  \left( v^2+16 \right) -\frac{1}{2}v\ln(v)  \nonumber \\
&&\left.  -i{\rm Li}_2 \left( -\frac{iv}{4} \right) +i{\rm Li}_2 \left( \frac{iv}{4} \right)\right]
\label{eq:alpha_ff_2}
\end{eqnarray}
where ${\rm Li}_2$ is the  the polylogarithm of index 2.
This quantity tends to a constant  at large bias:
\begin{equation}
\alpha(v\to\infty)/T_B =2.
\end{equation}
And at low bias we have:
\begin{equation}
  \frac{\pi}{T_B}\alpha(v\to0)	 =  \left( \frac{5}{288}-\frac{1}{48}\ln(v/4) \right) v^3+\mathcal{O} \left( v^5\right).
\end{equation}

\bibliography{irlm.bib}{}

\begin{thebibliography}{36}%
\makeatletter
\providecommand \@ifxundefined [1]{%
 \@ifx{#1\undefined}
}%
\providecommand \@ifnum [1]{%
 \ifnum #1\expandafter \@firstoftwo
 \else \expandafter \@secondoftwo
 \fi
}%
\providecommand \@ifx [1]{%
 \ifx #1\expandafter \@firstoftwo
 \else \expandafter \@secondoftwo
 \fi
}%
\providecommand \natexlab [1]{#1}%
\providecommand \enquote  [1]{``#1''}%
\providecommand \bibnamefont  [1]{#1}%
\providecommand \bibfnamefont [1]{#1}%
\providecommand \citenamefont [1]{#1}%
\providecommand \href@noop [0]{\@secondoftwo}%
\providecommand \href [0]{\begingroup \@sanitize@url \@href}%
\providecommand \@href[1]{\@@startlink{#1}\@@href}%
\providecommand \@@href[1]{\endgroup#1\@@endlink}%
\providecommand \@sanitize@url [0]{\catcode `\\12\catcode `\$12\catcode
  `\&12\catcode `\#12\catcode `\^12\catcode `\_12\catcode `\%12\relax}%
\providecommand \@@startlink[1]{}%
\providecommand \@@endlink[0]{}%
\providecommand \url  [0]{\begingroup\@sanitize@url \@url }%
\providecommand \@url [1]{\endgroup\@href {#1}{\urlprefix }}%
\providecommand \urlprefix  [0]{URL }%
\providecommand \Eprint [0]{\href }%
\providecommand \doibase [0]{http://dx.doi.org/}%
\providecommand \selectlanguage [0]{\@gobble}%
\providecommand \bibinfo  [0]{\@secondoftwo}%
\providecommand \bibfield  [0]{\@secondoftwo}%
\providecommand \translation [1]{[#1]}%
\providecommand \BibitemOpen [0]{}%
\providecommand \bibitemStop [0]{}%
\providecommand \bibitemNoStop [0]{.\EOS\space}%
\providecommand \EOS [0]{\spacefactor3000\relax}%
\providecommand \BibitemShut  [1]{\csname bibitem#1\endcsname}%
\let\auto@bib@innerbib\@empty
\bibitem [{\citenamefont {Polkovnikov}\ \emph {et~al.}(2011)\citenamefont
  {Polkovnikov}, \citenamefont {Sengupta}, \citenamefont {Silva},\ and\
  \citenamefont {Vengalattore}}]{Polkovnikov_RMP}%
  \BibitemOpen
  \bibfield  {author} {\bibinfo {author} {\bibfnamefont {A.}~\bibnamefont
  {Polkovnikov}}, \bibinfo {author} {\bibfnamefont {K.}~\bibnamefont
  {Sengupta}}, \bibinfo {author} {\bibfnamefont {A.}~\bibnamefont {Silva}}, \
  and\ \bibinfo {author} {\bibfnamefont {M.}~\bibnamefont {Vengalattore}},\
  }\href {\doibase 10.1103/RevModPhys.83.863} {\bibfield  {journal} {\bibinfo
  {journal} {Rev. Mod. Phys.}\ }\textbf {\bibinfo {volume} {83}},\ \bibinfo
  {pages} {863} (\bibinfo {year} {2011})}\BibitemShut {NoStop}%
\bibitem [{\citenamefont {Eisert}, \citenamefont {Friesdorf},\ and\
  \citenamefont {Gogolin}(2011)}]{eisert_quantum_2015}%
  \BibitemOpen
  \bibfield  {author} {\bibinfo {author} {\bibfnamefont {J.}~\bibnamefont
  {Eisert}}, \bibinfo {author} {\bibfnamefont {M.}~\bibnamefont {Friesdorf}}, \
  and\ \bibinfo {author} {\bibfnamefont {C.}~\bibnamefont {Gogolin}},\ }\href
  {\doibase 10.1038/nphys3215} {\bibfield  {journal} {\bibinfo  {journal} {Nat.
  Phys.}\ }\textbf {\bibinfo {volume} {11}},\ \bibinfo {pages} {124} (\bibinfo
  {year} {2011})}\BibitemShut {NoStop}%
\bibitem [{\citenamefont {Schollwöck}(2011)}]{schollwock_density-matrix_2011}%
  \BibitemOpen
  \bibfield  {author} {\bibinfo {author} {\bibfnamefont {U.}~\bibnamefont
  {Schollwöck}},\ }\href {\doibase 10.1016/j.aop.2010.09.012} {\bibfield
  {journal} {\bibinfo  {journal} {Ann. Phys.ics}\ }\textbf {\bibinfo {volume}
  {326}},\ \bibinfo {pages} {96} (\bibinfo {year} {2011})}\BibitemShut
  {NoStop}%
\bibitem [{\citenamefont {Hewson}(1993)}]{hewson_kondo_1993}%
  \BibitemOpen
  \bibfield  {author} {\bibinfo {author} {\bibfnamefont {A.~C.}\ \bibnamefont
  {Hewson}},\ }\href {https://doi.org/10.1017/CBO9780511470752} {\emph
  {\bibinfo {title} {The {Kondo} {Problem} to {Heavy} {Fermions}}}}\ (\bibinfo
  {publisher} {Cambridge University Press},\ \bibinfo {year}
  {1993})\BibitemShut {NoStop}%
\bibitem [{\citenamefont {Wiegmann}\ and\ \citenamefont
  {Finkel'shtein}(1978)}]{wiegmann_resonant-level_1978}%
  \BibitemOpen
  \bibfield  {author} {\bibinfo {author} {\bibfnamefont {P.~B.}\ \bibnamefont
  {Wiegmann}}\ and\ \bibinfo {author} {\bibfnamefont {A.~M.}\ \bibnamefont
  {Finkel'shtein}},\ }\href
  {http://www.jetp.ac.ru/cgi-bin/e/index/e/48/1/p102?a=list} {\bibfield
  {journal} {\bibinfo  {journal} {Sov. Phys. JETP}\ }\textbf {\bibinfo {volume}
  {48}},\ \bibinfo {pages} {102} (\bibinfo {year} {1978})}\BibitemShut
  {NoStop}%
\bibitem [{\citenamefont {Filyov}\ and\ \citenamefont
  {Wiegmann}(1980)}]{filyov_method_1980}%
  \BibitemOpen
  \bibfield  {author} {\bibinfo {author} {\bibfnamefont {V.~M.}\ \bibnamefont
  {Filyov}}\ and\ \bibinfo {author} {\bibfnamefont {P.~B.}\ \bibnamefont
  {Wiegmann}},\ }\href {\doibase 10.1016/0375-9601(80)90494-6} {\bibfield
  {journal} {\bibinfo  {journal} {Phys. Lett. A}\ }\textbf {\bibinfo {volume}
  {76}},\ \bibinfo {pages} {283} (\bibinfo {year} {1980})}\BibitemShut
  {NoStop}%
\bibitem [{\citenamefont {Boulat}, \citenamefont {Saleur},\ and\ \citenamefont
  {Schmitteckert}(2008)}]{boulat_twofold_2008}%
  \BibitemOpen
  \bibfield  {author} {\bibinfo {author} {\bibfnamefont {E.}~\bibnamefont
  {Boulat}}, \bibinfo {author} {\bibfnamefont {H.}~\bibnamefont {Saleur}}, \
  and\ \bibinfo {author} {\bibfnamefont {P.}~\bibnamefont {Schmitteckert}},\
  }\href {\doibase 10.1103/PhysRevLett.101.140601} {\bibfield  {journal}
  {\bibinfo  {journal} {Phys. Rev. Lett.}\ }\textbf {\bibinfo {volume} {101}},\
  \bibinfo {pages} {140601} (\bibinfo {year} {2008})}\BibitemShut {NoStop}%
\bibitem [{\citenamefont {Fendley}, \citenamefont {Ludwig},\ and\ \citenamefont
  {Saleur}(1995{\natexlab{a}})}]{fendley_exact_1995}%
  \BibitemOpen
  \bibfield  {author} {\bibinfo {author} {\bibfnamefont {P.}~\bibnamefont
  {Fendley}}, \bibinfo {author} {\bibfnamefont {A.~W.~W.}\ \bibnamefont
  {Ludwig}}, \ and\ \bibinfo {author} {\bibfnamefont {H.}~\bibnamefont
  {Saleur}},\ }\href {\doibase 10.1103/PhysRevLett.74.3005} {\bibfield
  {journal} {\bibinfo  {journal} {Phys. Rev. Lett.}\ }\textbf {\bibinfo
  {volume} {74}},\ \bibinfo {pages} {3005} (\bibinfo {year}
  {1995}{\natexlab{a}})}\BibitemShut {NoStop}%
\bibitem [{\citenamefont {Fendley}, \citenamefont {Ludwig},\ and\ \citenamefont
  {Saleur}(1995{\natexlab{b}})}]{fendley_exact_1995B}%
  \BibitemOpen
  \bibfield  {author} {\bibinfo {author} {\bibfnamefont {P.}~\bibnamefont
  {Fendley}}, \bibinfo {author} {\bibfnamefont {A.~W.~W.}\ \bibnamefont
  {Ludwig}}, \ and\ \bibinfo {author} {\bibfnamefont {H.}~\bibnamefont
  {Saleur}},\ }\href {\doibase 10.1103/PhysRevB.52.8934} {\bibfield  {journal}
  {\bibinfo  {journal} {Phys. Rev. B}\ }\textbf {\bibinfo {volume} {52}},\
  \bibinfo {pages} {8934} (\bibinfo {year} {1995}{\natexlab{b}})}\BibitemShut
  {NoStop}%
\bibitem [{\citenamefont {Mehta}\ and\ \citenamefont
  {Andrei}(2006)}]{mehta_nonequilibrium_2006}%
  \BibitemOpen
  \bibfield  {author} {\bibinfo {author} {\bibfnamefont {P.}~\bibnamefont
  {Mehta}}\ and\ \bibinfo {author} {\bibfnamefont {N.}~\bibnamefont {Andrei}},\
  }\href {\doibase 10.1103/PhysRevLett.96.216802} {\bibfield  {journal}
  {\bibinfo  {journal} {Phys. Rev. Lett.}\ }\textbf {\bibinfo {volume} {96}},\
  \bibinfo {pages} {216802} (\bibinfo {year} {2006})}\BibitemShut {NoStop}%
\bibitem [{\citenamefont {Gobert}\ \emph {et~al.}(2005)\citenamefont {Gobert},
  \citenamefont {Kollath}, \citenamefont {Schollwöck},\ and\ \citenamefont
  {Schütz}}]{gobert_real-time_2005}%
  \BibitemOpen
  \bibfield  {author} {\bibinfo {author} {\bibfnamefont {D.}~\bibnamefont
  {Gobert}}, \bibinfo {author} {\bibfnamefont {C.}~\bibnamefont {Kollath}},
  \bibinfo {author} {\bibfnamefont {U.}~\bibnamefont {Schollwöck}}, \ and\
  \bibinfo {author} {\bibfnamefont {G.}~\bibnamefont {Schütz}},\ }\href
  {\doibase 10.1103/PhysRevE.71.036102} {\bibfield  {journal} {\bibinfo
  {journal} {Phys. Rev. E}\ }\textbf {\bibinfo {volume} {71}},\ \bibinfo
  {pages} {036102} (\bibinfo {year} {2005})}\BibitemShut {NoStop}%
\bibitem [{\citenamefont {Sabetta}\ and\ \citenamefont
  {Misguich}(2013)}]{sabetta_nonequilibrium_2013}%
  \BibitemOpen
  \bibfield  {author} {\bibinfo {author} {\bibfnamefont {T.}~\bibnamefont
  {Sabetta}}\ and\ \bibinfo {author} {\bibfnamefont {G.}~\bibnamefont
  {Misguich}},\ }\href {\doibase 10.1103/PhysRevB.88.245114} {\bibfield
  {journal} {\bibinfo  {journal} {Phys. Rev. B}\ }\textbf {\bibinfo {volume}
  {88}},\ \bibinfo {pages} {245114} (\bibinfo {year} {2013})}\BibitemShut
  {NoStop}%
\bibitem [{\citenamefont
  {Schmitteckert}(2004)}]{schmitteckert_nonequilibrium_2004}%
  \BibitemOpen
  \bibfield  {author} {\bibinfo {author} {\bibfnamefont {P.}~\bibnamefont
  {Schmitteckert}},\ }\href {\doibase 10.1103/PhysRevB.70.121302} {\bibfield
  {journal} {\bibinfo  {journal} {Phys. Rev. B}\ }\textbf {\bibinfo {volume}
  {70}},\ \bibinfo {pages} {121302} (\bibinfo {year} {2004})}\BibitemShut
  {NoStop}%
\bibitem [{\citenamefont {Schneider}\ and\ \citenamefont
  {Schmitteckert}(2006)}]{schneider_conductance_2006}%
  \BibitemOpen
  \bibfield  {author} {\bibinfo {author} {\bibfnamefont {G.}~\bibnamefont
  {Schneider}}\ and\ \bibinfo {author} {\bibfnamefont {P.}~\bibnamefont
  {Schmitteckert}},\ }\href {http://arxiv.org/abs/cond-mat/0601389} {\bibfield
  {journal} {\bibinfo  {journal} {arXiv:cond-mat/0601389}\ } (\bibinfo {year}
  {2006})}\BibitemShut {NoStop}%
\bibitem [{\citenamefont {Vidal}(2004)}]{vidal_efficient_2004}%
  \BibitemOpen
  \bibfield  {author} {\bibinfo {author} {\bibfnamefont {G.}~\bibnamefont
  {Vidal}},\ }\href {\doibase 10.1103/PhysRevLett.93.040502} {\bibfield
  {journal} {\bibinfo  {journal} {Phys. Rev. Lett.}\ }\textbf {\bibinfo
  {volume} {93}},\ \bibinfo {pages} {040502} (\bibinfo {year}
  {2004})}\BibitemShut {NoStop}%
\bibitem [{\citenamefont {White}\ and\ \citenamefont
  {Feiguin}(2004)}]{white_real-time_2004}%
  \BibitemOpen
  \bibfield  {author} {\bibinfo {author} {\bibfnamefont {S.~R.}\ \bibnamefont
  {White}}\ and\ \bibinfo {author} {\bibfnamefont {A.~E.}\ \bibnamefont
  {Feiguin}},\ }\href {\doibase 10.1103/PhysRevLett.93.076401} {\bibfield
  {journal} {\bibinfo  {journal} {Phys. Rev. Lett.}\ }\textbf {\bibinfo
  {volume} {93}},\ \bibinfo {pages} {076401} (\bibinfo {year}
  {2004})}\BibitemShut {NoStop}%
\bibitem [{\citenamefont {Daley}\ \emph {et~al.}(2004)\citenamefont {Daley},
  \citenamefont {Kollath}, \citenamefont {Schollwöck},\ and\ \citenamefont
  {Vidal}}]{daley_time-dependent_2004}%
  \BibitemOpen
  \bibfield  {author} {\bibinfo {author} {\bibfnamefont {A.~J.}\ \bibnamefont
  {Daley}}, \bibinfo {author} {\bibfnamefont {C.}~\bibnamefont {Kollath}},
  \bibinfo {author} {\bibfnamefont {U.}~\bibnamefont {Schollwöck}}, \ and\
  \bibinfo {author} {\bibfnamefont {G.}~\bibnamefont {Vidal}},\ }\href
  {\doibase 10.1088/1742-5468/2004/04/P04005} {\bibfield  {journal} {\bibinfo
  {journal} {J. Stat. Mech.}\ }\textbf {\bibinfo {volume} {2004}},\ \bibinfo
  {pages} {P04005} (\bibinfo {year} {2004})}\BibitemShut {NoStop}%
\bibitem [{\citenamefont {Laflorencie}(2016)}]{laflorencie_quantum_2016}%
  \BibitemOpen
  \bibfield  {author} {\bibinfo {author} {\bibfnamefont {N.}~\bibnamefont
  {Laflorencie}},\ }\href {\doibase 10.1016/j.physrep.2016.06.008} {\bibfield
  {journal} {\bibinfo  {journal} {Phys. Rep.}\ }\textbf {\bibinfo {volume}
  {646}},\ \bibinfo {pages} {1} (\bibinfo {year} {2016})}\BibitemShut {NoStop}%
\bibitem [{ite(n 20)}]{itensor}%
  \BibitemOpen
  \href {http://itensor.org} {\bibfield  {journal} {\bibinfo  {journal}
  {ITensor Library, \href{http://itensor.org}{http://itensor.org}}\ } (\bibinfo
  {year} {version 2.0})}\BibitemShut {NoStop}%
\bibitem [{\citenamefont {Zaletel}\ \emph {et~al.}(2015)\citenamefont
  {Zaletel}, \citenamefont {Mong}, \citenamefont {Karrasch}, \citenamefont
  {Moore},\ and\ \citenamefont {Pollmann}}]{zaletel_time-evolving_2015}%
  \BibitemOpen
  \bibfield  {author} {\bibinfo {author} {\bibfnamefont {M.~P.}\ \bibnamefont
  {Zaletel}}, \bibinfo {author} {\bibfnamefont {R.~S.~K.}\ \bibnamefont
  {Mong}}, \bibinfo {author} {\bibfnamefont {C.}~\bibnamefont {Karrasch}},
  \bibinfo {author} {\bibfnamefont {J.~E.}\ \bibnamefont {Moore}}, \ and\
  \bibinfo {author} {\bibfnamefont {F.}~\bibnamefont {Pollmann}},\ }\href
  {\doibase 10.1103/PhysRevB.91.165112} {\bibfield  {journal} {\bibinfo
  {journal} {Phys. Rev. B}\ }\textbf {\bibinfo {volume} {91}},\ \bibinfo
  {pages} {165112} (\bibinfo {year} {2015})}\BibitemShut {NoStop}%
\bibitem [{\citenamefont {Eisler}\ and\ \citenamefont
  {Peschel}(2012)}]{eisler_entanglement_2012}%
  \BibitemOpen
  \bibfield  {author} {\bibinfo {author} {\bibfnamefont {V.}~\bibnamefont
  {Eisler}}\ and\ \bibinfo {author} {\bibfnamefont {I.}~\bibnamefont
  {Peschel}},\ }\href {\doibase 10.1209/0295-5075/99/20001} {\bibfield
  {journal} {\bibinfo  {journal} {EPL}\ }\textbf {\bibinfo {volume} {99}},\
  \bibinfo {pages} {20001} (\bibinfo {year} {2012})}\BibitemShut {NoStop}%
\bibitem [{\citenamefont {Kennes}, \citenamefont {Meden},\ and\ \citenamefont
  {Vasseur}(2014)}]{kennes_universal_2014}%
  \BibitemOpen
  \bibfield  {author} {\bibinfo {author} {\bibfnamefont {D.~M.}\ \bibnamefont
  {Kennes}}, \bibinfo {author} {\bibfnamefont {V.}~\bibnamefont {Meden}}, \
  and\ \bibinfo {author} {\bibfnamefont {R.}~\bibnamefont {Vasseur}},\ }\href
  {\doibase 10.1103/PhysRevB.90.115101} {\bibfield  {journal} {\bibinfo
  {journal} {Phys. Rev. B}\ }\textbf {\bibinfo {volume} {90}},\ \bibinfo
  {pages} {115101} (\bibinfo {year} {2014})}\BibitemShut {NoStop}%
\bibitem [{\citenamefont {Vasseur}\ and\ \citenamefont
  {Saleur}(2017)}]{vasseur_universal_2017}%
  \BibitemOpen
  \bibfield  {author} {\bibinfo {author} {\bibfnamefont {R.}~\bibnamefont
  {Vasseur}}\ and\ \bibinfo {author} {\bibfnamefont {H.}~\bibnamefont
  {Saleur}},\ }\href {\doibase 10.21468/SciPostPhys.3.1.001} {\bibfield
  {journal} {\bibinfo  {journal} {SciPost Physics}\ }\textbf {\bibinfo {volume}
  {3}},\ \bibinfo {pages} {001} (\bibinfo {year} {2017})}\BibitemShut {NoStop}%
\bibitem [{\citenamefont {Calabrese}\ and\ \citenamefont
  {Cardy}(2007)}]{calabrese_entanglement_2007}%
  \BibitemOpen
  \bibfield  {author} {\bibinfo {author} {\bibfnamefont {P.}~\bibnamefont
  {Calabrese}}\ and\ \bibinfo {author} {\bibfnamefont {J.}~\bibnamefont
  {Cardy}},\ }\href {\doibase 10.1088/1742-5468/2007/10/P10004} {\bibfield
  {journal} {\bibinfo  {journal} {J. Stat. Mech.}\ }\textbf {\bibinfo {volume}
  {2007}},\ \bibinfo {pages} {P10004} (\bibinfo {year} {2007})}\BibitemShut
  {NoStop}%
\bibitem [{\citenamefont {Stéphan}\ and\ \citenamefont
  {Dubail}(2011)}]{stephan_local_2011}%
  \BibitemOpen
  \bibfield  {author} {\bibinfo {author} {\bibfnamefont {J.-M.}\ \bibnamefont
  {Stéphan}}\ and\ \bibinfo {author} {\bibfnamefont {J.}~\bibnamefont
  {Dubail}},\ }\href {\doibase 10.1088/1742-5468/2011/08/P08019} {\bibfield
  {journal} {\bibinfo  {journal} {J. Stat. Mech.}\ }\textbf {\bibinfo {volume}
  {2011}},\ \bibinfo {pages} {P08019} (\bibinfo {year} {2011})}\BibitemShut
  {NoStop}%
\bibitem [{\citenamefont {Klich}\ and\ \citenamefont
  {Levitov}(2009)}]{klich_quantum_2009}%
  \BibitemOpen
  \bibfield  {author} {\bibinfo {author} {\bibfnamefont {I.}~\bibnamefont
  {Klich}}\ and\ \bibinfo {author} {\bibfnamefont {L.}~\bibnamefont
  {Levitov}},\ }\href {\doibase 10.1103/PhysRevLett.102.100502} {\bibfield
  {journal} {\bibinfo  {journal} {Phys. Rev. Lett.}\ }\textbf {\bibinfo
  {volume} {102}},\ \bibinfo {pages} {100502} (\bibinfo {year}
  {2009})}\BibitemShut {NoStop}%
\bibitem [{\citenamefont {Song}\ \emph {et~al.}(2012)\citenamefont {Song},
  \citenamefont {Rachel}, \citenamefont {Flindt}, \citenamefont {Klich},
  \citenamefont {Laflorencie},\ and\ \citenamefont
  {Le~Hur}}]{song_bipartite_2012}%
  \BibitemOpen
  \bibfield  {author} {\bibinfo {author} {\bibfnamefont {H.~F.}\ \bibnamefont
  {Song}}, \bibinfo {author} {\bibfnamefont {S.}~\bibnamefont {Rachel}},
  \bibinfo {author} {\bibfnamefont {C.}~\bibnamefont {Flindt}}, \bibinfo
  {author} {\bibfnamefont {I.}~\bibnamefont {Klich}}, \bibinfo {author}
  {\bibfnamefont {N.}~\bibnamefont {Laflorencie}}, \ and\ \bibinfo {author}
  {\bibfnamefont {K.}~\bibnamefont {Le~Hur}},\ }\href {\doibase
  10.1103/PhysRevB.85.035409} {\bibfield  {journal} {\bibinfo  {journal} {Phys.
  Rev. B}\ }\textbf {\bibinfo {volume} {85}},\ \bibinfo {pages} {035409}
  (\bibinfo {year} {2012})}\BibitemShut {NoStop}%
\bibitem [{\citenamefont {Song}\ \emph {et~al.}(2011)\citenamefont {Song},
  \citenamefont {Flindt}, \citenamefont {Rachel}, \citenamefont {Klich},\ and\
  \citenamefont {Le~Hur}}]{song_entanglement_2011}%
  \BibitemOpen
  \bibfield  {author} {\bibinfo {author} {\bibfnamefont {H.~F.}\ \bibnamefont
  {Song}}, \bibinfo {author} {\bibfnamefont {C.}~\bibnamefont {Flindt}},
  \bibinfo {author} {\bibfnamefont {S.}~\bibnamefont {Rachel}}, \bibinfo
  {author} {\bibfnamefont {I.}~\bibnamefont {Klich}}, \ and\ \bibinfo {author}
  {\bibfnamefont {K.}~\bibnamefont {Le~Hur}},\ }\href {\doibase
  10.1103/PhysRevB.83.161408} {\bibfield  {journal} {\bibinfo  {journal} {Phys.
  Rev. B}\ }\textbf {\bibinfo {volume} {83}},\ \bibinfo {pages} {161408}
  (\bibinfo {year} {2011})}\BibitemShut {NoStop}%
\bibitem [{\citenamefont {Karrasch}\ \emph {et~al.}(2010)\citenamefont
  {Karrasch}, \citenamefont {Pletyukhov}, \citenamefont {Borda},\ and\
  \citenamefont {Meden}}]{karrasch_functional_2010}%
  \BibitemOpen
  \bibfield  {author} {\bibinfo {author} {\bibfnamefont {C.}~\bibnamefont
  {Karrasch}}, \bibinfo {author} {\bibfnamefont {M.}~\bibnamefont
  {Pletyukhov}}, \bibinfo {author} {\bibfnamefont {L.}~\bibnamefont {Borda}}, \
  and\ \bibinfo {author} {\bibfnamefont {V.}~\bibnamefont {Meden}},\ }\href
  {\doibase 10.1103/PhysRevB.81.125122} {\bibfield  {journal} {\bibinfo
  {journal} {Phys. Rev. B}\ }\textbf {\bibinfo {volume} {81}},\ \bibinfo
  {pages} {125122} (\bibinfo {year} {2010})}\BibitemShut {NoStop}%
\bibitem [{\citenamefont {Vinkler-Aviv}, \citenamefont {Schiller},\ and\
  \citenamefont {Anders}(2014)}]{vinkler-aviv_thermal_2014}%
  \BibitemOpen
  \bibfield  {author} {\bibinfo {author} {\bibfnamefont {Y.}~\bibnamefont
  {Vinkler-Aviv}}, \bibinfo {author} {\bibfnamefont {A.}~\bibnamefont
  {Schiller}}, \ and\ \bibinfo {author} {\bibfnamefont {F.~B.}\ \bibnamefont
  {Anders}},\ }\href {\doibase 10.1103/PhysRevB.90.155110} {\bibfield
  {journal} {\bibinfo  {journal} {Phys. Rev. B}\ }\textbf {\bibinfo {volume}
  {90}},\ \bibinfo {pages} {155110} (\bibinfo {year} {2014})}\BibitemShut
  {NoStop}%
\bibitem [{\citenamefont {Freton}, \citenamefont {Boulat},\ and\ \citenamefont
  {Saleur}(2013)}]{freton_infrared_2013}%
  \BibitemOpen
  \bibfield  {author} {\bibinfo {author} {\bibfnamefont {L.}~\bibnamefont
  {Freton}}, \bibinfo {author} {\bibfnamefont {E.}~\bibnamefont {Boulat}}, \
  and\ \bibinfo {author} {\bibfnamefont {H.}~\bibnamefont {Saleur}},\ }\href
  {\doibase 10.1016/j.nuclphysb.2013.05.015} {\bibfield  {journal} {\bibinfo
  {journal} {Nucl. Phys. B}\ }\textbf {\bibinfo {volume} {874}},\ \bibinfo
  {pages} {279} (\bibinfo {year} {2013})}\BibitemShut {NoStop}%
\bibitem [{\citenamefont {Branschädel}\ \emph {et~al.}(2010)\citenamefont
  {Branschädel}, \citenamefont {Boulat}, \citenamefont {Saleur},\ and\
  \citenamefont {Schmitteckert}}]{branschadel_numerical_2010}%
  \BibitemOpen
  \bibfield  {author} {\bibinfo {author} {\bibfnamefont {A.}~\bibnamefont
  {Branschädel}}, \bibinfo {author} {\bibfnamefont {E.}~\bibnamefont
  {Boulat}}, \bibinfo {author} {\bibfnamefont {H.}~\bibnamefont {Saleur}}, \
  and\ \bibinfo {author} {\bibfnamefont {P.}~\bibnamefont {Schmitteckert}},\
  }\href {\doibase 10.1103/PhysRevB.82.205414} {\bibfield  {journal} {\bibinfo
  {journal} {Phys. Rev. B}\ }\textbf {\bibinfo {volume} {82}},\ \bibinfo
  {pages} {205414} (\bibinfo {year} {2010})}\BibitemShut {NoStop}%
\bibitem [{\citenamefont {Landauer}(1957)}]{landauer_spatial_1957}%
  \BibitemOpen
  \bibfield  {author} {\bibinfo {author} {\bibfnamefont {R.}~\bibnamefont
  {Landauer}},\ }\href {\doibase 10.1147/rd.13.0223} {\bibfield  {journal}
  {\bibinfo  {journal} {IBM J. Res. Dev.}\ }\textbf {\bibinfo {volume} {1}},\
  \bibinfo {pages} {223} (\bibinfo {year} {1957})}\BibitemShut {NoStop}%
\bibitem [{\citenamefont {Imry}\ and\ \citenamefont
  {Landauer}(1999)}]{imry_conductance_1999}%
  \BibitemOpen
  \bibfield  {author} {\bibinfo {author} {\bibfnamefont {Y.}~\bibnamefont
  {Imry}}\ and\ \bibinfo {author} {\bibfnamefont {R.}~\bibnamefont
  {Landauer}},\ }\href {\doibase 10.1103/RevModPhys.71.S306} {\bibfield
  {journal} {\bibinfo  {journal} {Rev. Mod. Phys.}\ }\textbf {\bibinfo {volume}
  {71}},\ \bibinfo {pages} {S306} (\bibinfo {year} {1999})}\BibitemShut
  {NoStop}%
\bibitem [{\citenamefont {Nazarov}\ and\ \citenamefont
  {Blanter}(2009)}]{nazarov_quantum_2009}%
  \BibitemOpen
  \bibfield  {author} {\bibinfo {author} {\bibfnamefont {Y.~V.}\ \bibnamefont
  {Nazarov}}\ and\ \bibinfo {author} {\bibfnamefont {Y.}~\bibnamefont
  {Blanter}},\ }\href
  {http://www.cambridge.org/fr/academic/subjects/physics/condensed-matter-physics-nanoscience-and-mesoscopic-physics/quantum-transport-introduction-nanoscience}
  {\emph {\bibinfo {title} {Quantum transport - Introduction to Nanoscience}}}\
  (\bibinfo  {publisher} {Cambridge University Press},\ \bibinfo {year}
  {2009})\BibitemShut {NoStop}%
\bibitem [{\citenamefont {Antal}, \citenamefont {Krapivsky},\ and\
  \citenamefont {Rákos}(2008)}]{antal_logarithmic_2008}%
  \BibitemOpen
  \bibfield  {author} {\bibinfo {author} {\bibfnamefont {T.}~\bibnamefont
  {Antal}}, \bibinfo {author} {\bibfnamefont {P.~L.}\ \bibnamefont
  {Krapivsky}}, \ and\ \bibinfo {author} {\bibfnamefont {A.}~\bibnamefont
  {Rákos}},\ }\href {\doibase 10.1103/PhysRevE.78.061115} {\bibfield
  {journal} {\bibinfo  {journal} {Phys. Rev. E}\ }\textbf {\bibinfo {volume}
  {78}},\ \bibinfo {pages} {061115} (\bibinfo {year} {2008})}\BibitemShut
  {NoStop}%
\end{thebibliography}%

\end{document}